\documentclass[pra, twocolumn, final, reprint, amsmath, amssymb, superscriptaddress, showpacs, a4paper, aps, 10pt, hidelinks, numbers, sort&compress, floats, balancelastpage]{revtex4-1}

\usepackage{graphicx, color} 
\usepackage[normalem]{ulem}          

\usepackage[colorlinks=true]{hyperref}
\hypersetup{pdfpagemode=None, linkcolor=black, citecolor=blue, urlcolor=blue}  

\usepackage[utf8]{inputenc}

\usepackage{multibib}

\newcommand{\ket}[1]{\ensuremath{\left|#1\right\rangle}}

\newcommand{\1}{\ensuremath{\left|1 \right\rangle}}
\newcommand{\2}{\ensuremath{\left|2 \right\rangle}}
\newcommand{\3}{\ensuremath{\left|3 \right\rangle}}

\definecolor{britishracinggreen}{rgb}{0.0, 0.26, 0.15}
\definecolor{bulgarianrose}{rgb}{0.28, 0.02, 0.03}
\definecolor{darkred}{rgb}{0.90,0,0}
\definecolor{darkgreen}{rgb}{0,0.60,.2}
\definecolor{darkblue}{rgb}{0,0,1}
\definecolor{orange}{cmyk}{0,0.6,0.8,0}
\definecolor{lightblue}{rgb}{0.3,0.5,1}
\definecolor{lightgreen}{rgb}{0.4,0.80,.4}

\newcommand{\Li}{$^{6}$Li }

\newcommand{\beq}{\begin{equation}}
\newcommand{\eeq}{\end{equation}}
\newcommand{\bei}{\begin{itemize}}
\newcommand{\eei}{\end{itemize}}
\newcommand{\ben}{\begin{enumerate}}
\newcommand{\een}{\end{enumerate}}

\newcommand{\bk}{{\bf k} }
\newcommand{\bp}{{\bf p} }
\newcommand{\bq}{{\bf q} }
\newcommand{\br}{{\bf r} }
\newcommand{\down}{\downarrow}
\newcommand{\up}{\uparrow}

\begin{document}

\title{Repulsive Fermi polarons in a resonant mixture of ultracold $^6$Li atoms}


\author{F. Scazza}
\email{scazza@lens.unifi.it}
\affiliation{Istituto Nazionale di Ottica del Consiglio Nazionale delle Ricerche (INO-CNR), 50019 Sesto Fiorentino, Italy}
\affiliation{\mbox{LENS and Dipartimento di Fisica e Astronomia, Universit\`{a} di Firenze, 50019 Sesto Fiorentino, Italy}}
\author{G. Valtolina}
\affiliation{Istituto Nazionale di Ottica del Consiglio Nazionale delle Ricerche (INO-CNR), 50019 Sesto Fiorentino, Italy}
\affiliation{\mbox{LENS and Dipartimento di Fisica e Astronomia, Universit\`{a} di Firenze, 50019 Sesto Fiorentino, Italy}}
\author{P. Massignan}
\affiliation{ICFO-Institut de Ciencies Fotoniques, The Barcelona Institute of Science and Technology, 08860 Castelldefels, 
Spain}
\author{A. Recati}
\affiliation{INO-CNR BEC Center and Dipartimento di Fisica, Universit\`a di Trento, 38123 Povo, Italy}
\affiliation{Ludwig-Maximilians-Universit\"at M\"unchen, 80333 M\"unchen, Germany}
\author{A. Amico}
\affiliation{\mbox{LENS and Dipartimento di Fisica e Astronomia, Universit\`{a} di Firenze, 50019 Sesto Fiorentino, Italy}}
\author{A.~Burchianti}
\affiliation{Istituto Nazionale di Ottica del Consiglio Nazionale delle Ricerche (INO-CNR), 50019 Sesto Fiorentino, Italy}
\affiliation{\mbox{LENS and Dipartimento di Fisica e Astronomia, Universit\`{a} di Firenze, 50019 Sesto Fiorentino, Italy}}
\author{C. Fort}
\affiliation{\mbox{LENS and Dipartimento di Fisica e Astronomia, Universit\`{a} di Firenze, 50019 Sesto Fiorentino, Italy}}
\author{M. Inguscio}
\affiliation{Istituto Nazionale di Ottica del Consiglio Nazionale delle Ricerche (INO-CNR), 50019 Sesto Fiorentino, Italy}
\affiliation{\mbox{LENS and Dipartimento di Fisica e Astronomia, Universit\`{a} di Firenze, 50019 Sesto Fiorentino, Italy}}
\author{M. Zaccanti}
\affiliation{Istituto Nazionale di Ottica del Consiglio Nazionale delle Ricerche (INO-CNR), 50019 Sesto Fiorentino, Italy}
\affiliation{\mbox{LENS and Dipartimento di Fisica e Astronomia, Universit\`{a} di Firenze, 50019 Sesto Fiorentino, Italy}}
\author{G. Roati}
\affiliation{Istituto Nazionale di Ottica del Consiglio Nazionale delle Ricerche (INO-CNR), 50019 Sesto Fiorentino, Italy}
\affiliation{\mbox{LENS and Dipartimento di Fisica e Astronomia, Universit\`{a} di Firenze, 50019 Sesto Fiorentino, Italy}}

%



\begin{abstract}
We employ radio-frequency spectroscopy to investigate a polarized spin-mixture of ultracold \Li atoms close to a broad Feshbach scattering resonance. Focusing on the regime of strong repulsive interactions, 
we observe well-defined coherent quasiparticles even for unitarity-limited interactions. We characterize the many-body system by extracting the key properties of repulsive Fermi polarons: the energy $E_+$, the effective mass $m^*$, the residue $Z$ and the decay rate $\Gamma$. 
Above a critical interaction, $E_+$ is found to exceed the Fermi energy of the bath while $m^*$ diverges and even turns negative, thereby indicating that the repulsive Fermi liquid state becomes energetically and thermodynamically unstable.
\end{abstract}

\maketitle

Landau's idea of mapping the behavior of impurity particles interacting with a complex environment into quasiparticle properties \cite{Landau1957} plays a fundamental role in physics and materials science, from helium liquids \cite{Bardeen1967} and colossal magnetoresistive materials \cite{Teresa1997, Millis1998} to polymers and proteins \cite{Deibel2010, Davydov1973}. 
In the field of ultracold gases, the impurity problem and the associated concept of polaron quasiparticle have attracted over the last decade a growing interest \cite{Radzihovsky2010, Chevy2010, Knap2012, Massignan2014}. Initiated with the investigation of polarized Fermi gases in the BEC-BCS crossover \cite{Lobo2006, Zwierlein2006, Partridge2006, Schirotzek2009, Nascimbene2009, Navon2010}, the study of polaron physics has been extended to mass-imbalanced \cite{Kohstall2012, Cetina2016}, low-dimensional fermionic systems \cite{Koschorreck2012}, and also to bosonic environments \cite{Catani2012, Hu2016, Jorgensen2016}.  
The polaron properties are fundamentally relevant for understanding the more complex scenario of partially-polarized and balanced Fermi mixtures: the impurity limit exhibits some of the critical points of the full phase diagram, whose topology we can thus learn about by investigating polarized systems \cite{Chevy2010, Navon2010}.

\enlargethispage{\baselineskip}
While researchers initially focused on attractive interactions \cite{Schirotzek2009, Nascimbene2009}, more recently they have explored novel quasiparticles associated with repulsive interactions: these repulsive polarons \cite{Cui2010, Pilati2010, Massignan2011, Schmidt2011, Massignan2013} 
are centrally important for realizing repulsive many-body states \cite{Duine2005, Cui2010, Pilati2010, Chang2011} and therein exploring itinerant ferromagnetism \cite{Jo2009, Sanner2012, Valtolina2016}. In particular, if the polaron energy exceeds the Fermi energy of the surrounding medium, a fully-ferromagnetic phase is favored against the paramagnetic Fermi liquid \cite{Cui2010, Pilati2010, Massignan2011, Massignan2013}. 
However, short-ranged strong repulsion always require an underlying weakly-bound molecular state, into which the system may rapidly decay \cite{Pekker2011, Sanner2012}, making the repulsive polaron an excited many-body state, whose theoretical and experimental investigation are challenging. 
In three dimensions, repulsive Fermi polarons have been first unveiled in a $^6$Li\,\,-\,$^{40}$K mixture at a comparatively narrow Feshbach resonance \cite{Kohstall2012}, but they lack observation in the universal, broad resonance case, for which the decay rate is expected to be the largest~\cite{Massignan2014}.

In this Letter we report on reverse radio-frequency (RF) spectroscopy \cite{Gupta2003, Regal2003a, Kohstall2012} experiments to unveil the existence and characterize the properties of repulsive polarons in a polarized Fermi mixture of lithium atoms, interacting at a broad Feshbach resonance. We obtain precise information about: (i) the energy $E_+$, (ii) the effective mass $m^*$ and (iii) the decay rate $\Gamma$ of repulsive polarons. Furthermore, 
we probe the coherence properties of these fermionic excitations and extract (iv) the quasiparticle residue $Z$.
Our findings imply that phase separation is energetically allowed above a critical value of repulsion, where $E_+$ is found to exceed the Fermi energy of the majority atoms \cite{Cui2010, Pilati2010, Massignan2011, Massignan2013}. We also observe a negative effective mass at strong coupling, which points to a thermodynamical instability of the repulsive polaron Fermi liquid \cite{Combescot2009, Trefzger2012}. 
Unexpectedly, the measured decay rate of the repulsive branch population at the critical point is less than $0.1\,E_+$, and never exceeds $0.2\,E_+$, demonstrating the existence of well-defined repulsive quasiparticles even for resonant interactions.

\begin{figure}[t!]
\begin{center}
\includegraphics[width= 0.93\columnwidth]{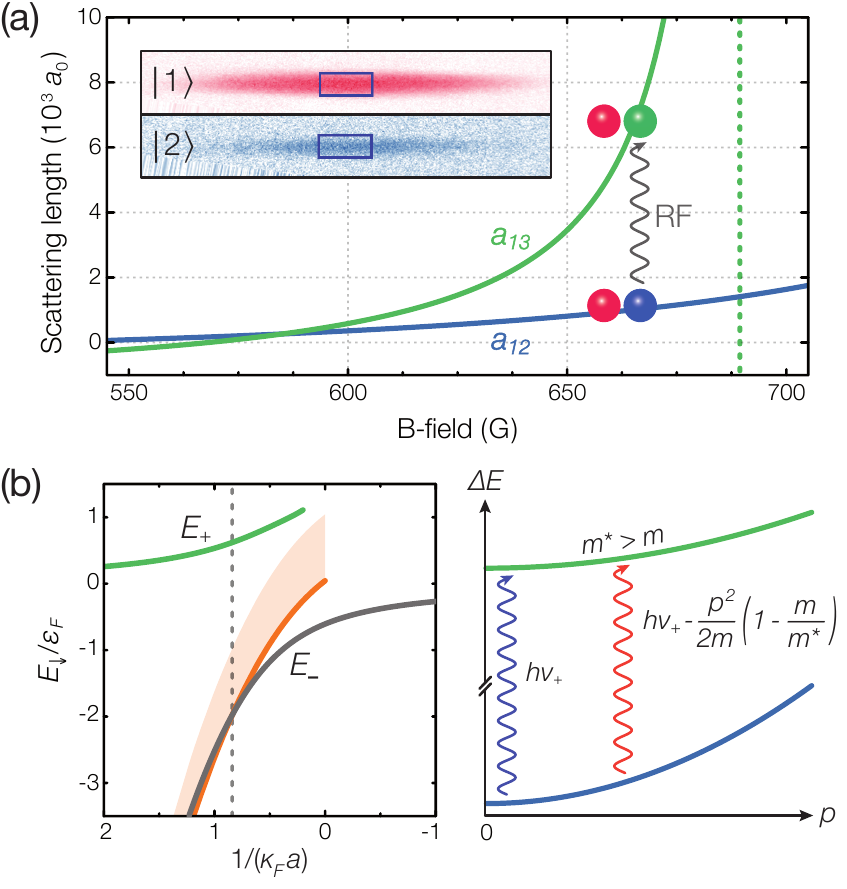}
\caption{ (a) Scattering properties of the initial 1-2 and final 1-3 state mixtures. The RF pulse pumps impurities from the weakly interacting into the resonant state. Inset: \textit{In situ} absorption images of the state \1 (red) and \2 (blue) atomic clouds for $x=0.10(1)$. The rectangles mark the central region where the spectroscopy signal is recorded. (b) Left: Energy $E_\downarrow$ landscape of a zero-momentum impurity interacting with a homogeneous Fermi gas. The shaded area denotes the dressed dimer continuum and the vertical dashed line marks the polaron/molecule crossing. Right: Sketch of the momentum dependence of the impurity resonance frequency. The blue (green) curve depicts the dispersion of bare (dressed) impurities.}
\label{Fig1}
\end{center}
\end{figure}

In our experiment, we initially produce a weakly interacting imbalanced mixture of \Li atoms in the two lowest Zeeman states, hereafter denoted as \1 and \2 respectively, held in a crossed optical dipole trap at a bias magnetic field of 300\,G \cite{SM}. The majority \1-component forms a highly degenerate Fermi gas with $N_1 \simeq 1.5 \times 10^5$\,atoms at $T/T_F = 0.10(2)$, where $E_F=k_B T_F\simeq h \times 9.5\,$kHz is the Fermi energy, and $k_B$ and $h$ respectively denote the Boltzmann and Planck constants. The state-\1 Fermi gas acts as a bath for the minority state-\2 impurities, whose concentration $x=N_2/N_1$ is finely adjusted between 0.05 and 0.4 \cite{SM}. 
To explore strong impurity-bath interactions, we exploit the third-to-lowest Zeeman state \3, and the tunability of the scattering lengths $a_{12}$ and $a_{13}$ enabled by two off-centered broad Feshbach resonances between the 1-2 and 1-3 spin combinations. Upon increasing the bias field to values between 600\,G and 700\,G, we resonantly enhance the 1-3 scattering while moderately increasing the comparatively weak 1-2 interactions (see Fig.~\ref{Fig1}(a)). 

\begin{figure}[t!]
\begin{center}
\includegraphics[width= \columnwidth]{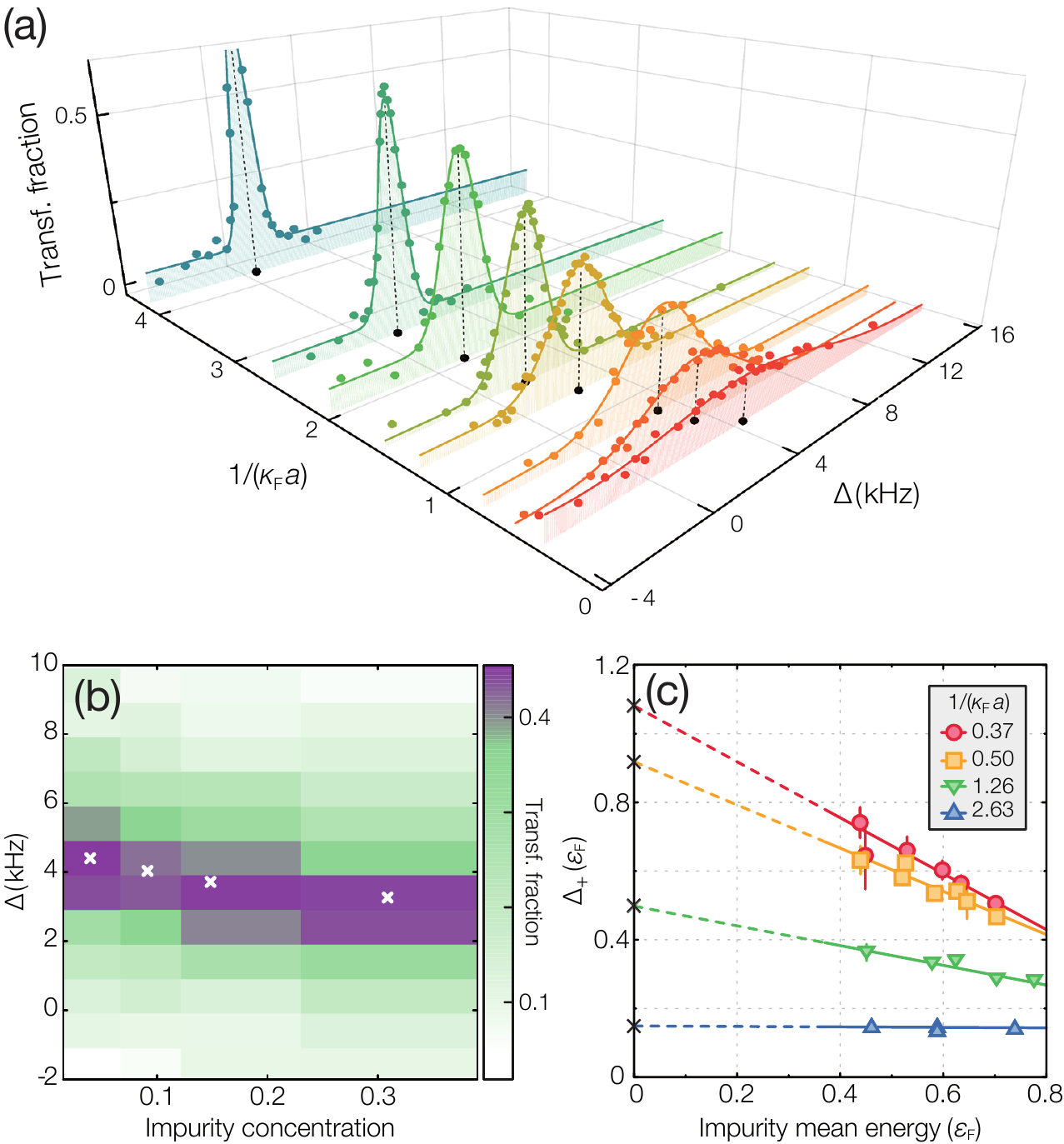}
\caption{
Examples of repulsive polaron spectral response recorded (a) at different $1/(\kappa_F a)$ values with concentration $x = 0.15(3)$, and (b) at different $x$ values with $\kappa_F a\simeq 2$. (c) Resonance position $\Delta_+$ as a function of $\bar{\varepsilon}$ for various $1/(\kappa_F a)$ values (see legend). The linear fits used to extract $E_+$ and $m/m^*$ are shown. Error bars denote the standard errors of the fitted $\Delta_+$.}
\label{Fig2}
\end{center}
\end{figure}

In order to probe the full excitation spectrum of the impurities, we employ reverse RF spectroscopy \cite{Gupta2003, Regal2003a, Kohstall2012}: we drive the \2 atoms on the $\2 \rightarrow \3$ transition to the resonantly interacting state, using a RF pulse with variable frequency $\nu$. 
Our spectroscopy signal is the transferred fraction $\mathcal{N}_3/(\mathcal{N}_2 + \mathcal{N}_3)$, where $\mathcal{N}_{i}$ is the number of $\ket{i}$ atoms contained in a centered region of size 70\,$\mu$m (30\,$\mu$m) along the axial (transverse) direction of the trap (see Fig.~\ref{Fig1}(a)). For each experimental run, the populations $\mathcal{N}_2$ and $\mathcal{N}_3$ are separately monitored by acquiring two consecutive \textit{in situ} absorption images delayed by 500\,$\mu$s. The transferred fraction is measured as a function of the RF detuning $\Delta = \nu - \nu_0$ from the frequency $\nu_0$ of the non-interacting RF transition, measured in absence of majority atoms. 
Extracting the signal from such a central region helps to reduce the effects of density inhomogeneity. The bath is characterized by effective Fermi energy $\varepsilon_F \simeq 0.74\,E_F$ and wavevector $\kappa_F \simeq 0.86\,k_F$, averaged over the \textit{in situ} density distribution of the state-\1 gas within the integration region \cite{SM}. The bath residual inhomogeneity quantified by a standard deviation $\Delta \kappa_F \sim 0.1\,\kappa_F$. 
From here on, interactions will be parametrized by $1/(\kappa_F a) \equiv 1/(\kappa_F a_{13})$.

Figure~\ref{Fig1}(b) illustrates the generic energy spectrum of a zero-momentum impurity in a Fermi sea 
in the mass-balanced and broad resonance case. 
Attractive and repulsive polarons appear as discrete levels, with monotonically increasing energies $E_+$ and $E_-$ as $1/(\kappa_F a)$ is decreased. Moreover, the repulsive polaron acquires an increasingly large width (not shown), owing to a non-zero probability to decay onto lower-lying states. These also include a broad continuum of molecular excitations of spectral width $\sim \varepsilon_F$, which arise from processes in which the impurity and any of the majority fermions are bound into a molecule.  
The attractive polaron enters the molecular continuum for $1/(\kappa_F a) \gtrsim 0.9$ \cite{Combescot2009, Prokofev2008}, beyond which a dressed molecule becomes energetically favored.
Reverse RF spectroscopy allows to entirely explore this energy landscape: besides a broad molecular state contribution, peaks in the RF signal centered at $\Delta_+ > 0$ ($\Delta_- <0$) are identified as the repulsive (attractive) polaron states, providing access to $E_+$ ($E_-$). 

Typical repulsive polaron spectra at various $1/(\kappa_F a)$ values, obtained using a 1\,ms-long rectangular pulse, i.e.~a $0.8\,\pi$-pulse for non-interacting impurity atoms, are displayed in Fig.~\ref{Fig2}(a). These are shown together with Gaussian fits employed to extract the resonance position $\Delta_+$. 
The RF shift $\Delta_+$ increases monotonically when increasing $\kappa_F a$, while the resonance progressively widens, owing mainly to collisional broadening in the final state \cite{Kohstall2012, Cetina2016, SM}. The resonance shift reflects the increase of polaron energy due to the repulsion between the impurities and the surrounding medium. 
However, the link between the measured $\Delta_+$ and the zero-momentum polaron energy $E_+$ is complicated by an observed strong dependence of $\Delta_+$ upon the impurity concentration $x$ (see Fig.~\ref{Fig2}(b)). 
This can in principle arise from two distinct effects. A first effect is associated to the different dispersions featured by the initial weakly-interacting impurity, characterized by the bare atomic mass $m$, and by the final quasiparticle with effective mass $m^*$ (see Fig.~\ref{Fig1}(b)).
Increasing $x$ from 0 to 1 at fixed $T \simeq 0.1\,T_F$, the mean motional energy per impurity in the region of interest grows non-linearly from $\bar{\varepsilon}\simeq 0.42\,\varepsilon_F$ to $\bar{\varepsilon}\simeq \varepsilon_F$, due to the increased Fermi pressure of the minority gas \cite{SM}. Since the RF driving transfers the impurities into final polaron states without modifying their momentum, we expect the measured resonance shift $\Delta_+$ at fixed $\kappa_F a$ to depend linearly upon the mean impurity energy $\bar{\varepsilon}$, with a negative slope directly reflecting the value of $m^*$ \cite{SM}:
\begin{equation}
	\Delta_+ = E_+ - (1-\frac{m}{m^*})\, \bar{\varepsilon}
	\label{Eq1}
\end{equation}
Our data indeed exhibit such a linear decrease of $\Delta_+$ with increasing $\bar{\varepsilon}$ (see Fig.~\ref{Fig2}(b)-(c)).

On the other hand, polaron-polaron effective interactions are expected, within an equilibrium Fermi liquid, to contribute with a positive resonance shift $\propto  x \sim \varepsilon^{3/2}$ \cite{Bardeen1967, Mora2010, Yu2012a, SM}, leading to a non-linear increase of $\Delta_+$ with $\bar{\varepsilon}$. 
Such trend is incompatible with the observed linear decrease. Furthermore, 
sizeable effective interactions would induce additional spectral broadening and decoherence \cite{Cetina2016} for increasing $x$, never exhibited by our data (see Fig.~\ref{Fig2}(b) and Ref.~\cite{SM}). 
Therefore, our measurements show no evidence of polaron interaction effects.
In light of this, for all explored values of $\kappa_F a$, we extract the polaron energy $E_+$ and effective mass $m^*$ by fitting our data with Eq.~\eqref{Eq1}. In determining these quantities, we have also taken into account the weak initial interaction energy of state-\2 impurities in the state-\1 medium (see Fig.~\ref{Fig1}(a)), and the associated tiny mass~\mbox{renormalization}~\cite{SM}.

\begin{figure}[b]
\begin{center}
\vspace*{-10pt}
\includegraphics[width=0.9\columnwidth]{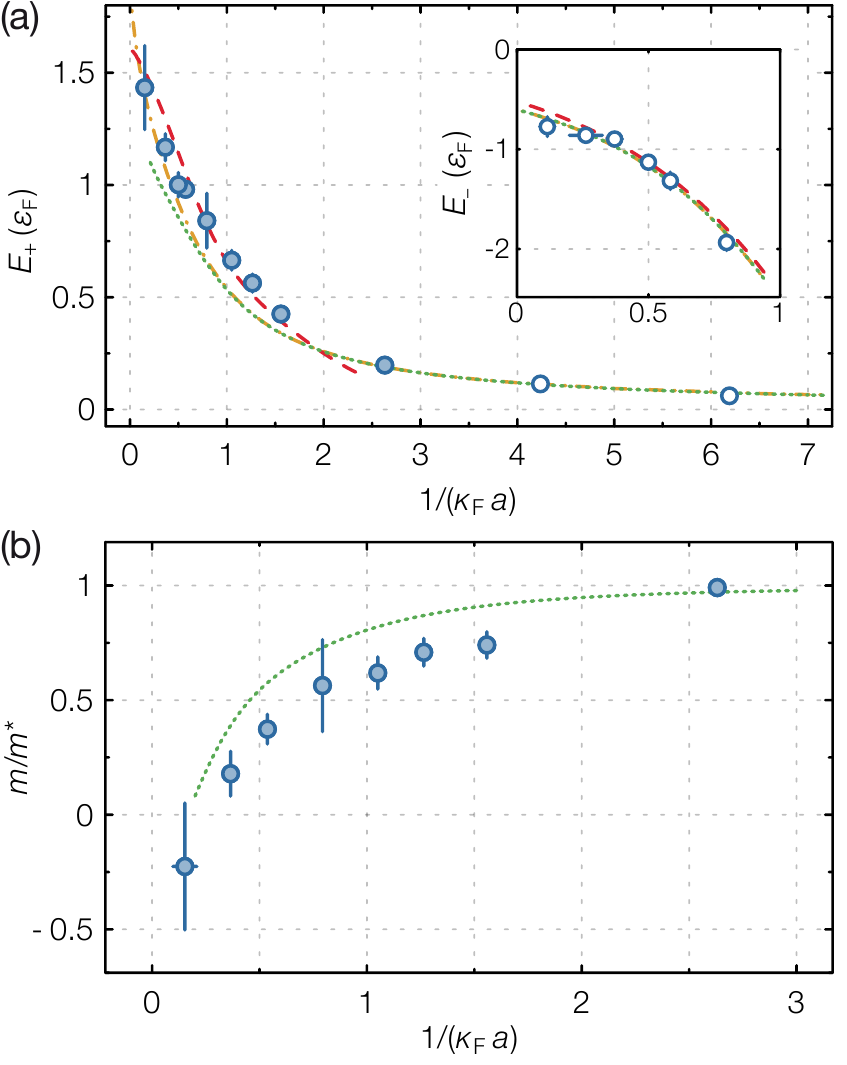}
\caption{(a) Zero-momentum repulsive polaron energy $E_+$ as a function of $1/(\kappa_F a)$ (symbols). Inset: attractive polaron energy $E_-$. Theory predictions from Ref.~\cite{Cui2010} (dot-dashed yellow line), Ref.~\cite{Massignan2011} (dotted green line) and Ref.~\cite{Schmidt2011} (dashed red line) are shown in both panels. Empty symbols denote points obtained by averaging measurements at different $\bar{\varepsilon}$ rather than by zero-energy extrapolation \cite{SM}. 
(b) Inverse effective mass $m/m^*$ of the repulsive polaron as a function of $1/\kappa_F a$ (symbols), together with theory predictions from Ref.~\cite{Massignan2011} (dotted green line). Error bars combine the linear fit parameter errors with the standard error of the mean  (s.e.m.) of binned data.} 
\label{Fig3}
\end{center}
\end{figure}

The determined behaviors of $E_+$ and $m/m^*$ are presented in Fig.~\ref{Fig3}(a) and~\ref{Fig3}(b), respectively. The polaron energy $E_+$ is found in good agreement with recent $T=0$ theoretical predictions based either on a variational model \cite{Cui2010}, on diagrammatic calculations within the ladder approximation \cite{Massignan2011, SM} or on the functional renormalization group \cite{Schmidt2011}, which in turn compare well to quantum Monte Carlo simulations (QMC) \cite{Pilati2010}.  
Importantly, for $1/(\kappa_F a) < 1/(\kappa_F a)_{c} = 1/1.7(2) \simeq 0.6(1)$, $E_+$ exceeds $\varepsilon_F$, indicating that the Fermi liquid of repulsive polarons becomes energetically disfavored against a phase-separated state~\cite{Cui2010, Pilati2010, Massignan2011}. This value of $(\kappa_F a)_{c}$ is larger than that recently reported for a balanced spin-mixture \cite{Valtolina2016}, consistently with QMC predictions~\cite{Pilati2010}.
In the inset of Fig.~\ref{Fig3}(a) we also present the attractive polaron energy $E_-$, extracted by fitting the resonances at $\Delta < 0$ in the spectra recorded at strong interactions (see \cite{SM} for further details). 
Here, we find excellent agreement with theories and previous experiments \cite{Prokofev2008, Schirotzek2009, Schmidt2011,Massignan2011}. 

The behavior of the repulsive polaron effective mass provides also important information:~the extracted $m/m^*$ strongly decreases for increasing $\kappa_F a$, until it becomes zero and eventually turns negative at very strong repulsion. This feature, never observed experimentally, has been previously pointed out in the context of attractive Fermi polarons \cite{Combescot2009, Trefzger2012}. There, a negative $m^*$ has been predicted for interaction strengths well beyond the polaron-molecule crossing, and interpreted as a signature for the attractive polaron being thermodynamically unstable against the dressed dimer. Similarly, the observation of $m^*<0$ at $\kappa_F a > \kappa_F a_c$ suggests a thermodynamic instability of the repulsive Fermi liquid. 
Overall, the experimental trend of $m^*$ is reasonably reproduced by the theory from Ref.~\cite{Massignan2011} (see line in Fig.~\ref{Fig3}(b)), which is however expected to underestimate $m^*$ since it includes only one-particle-hole excitations \cite{Combescot2009}. 

\begin{figure}[t!]
\begin{center}
\includegraphics[width=0.9\columnwidth]{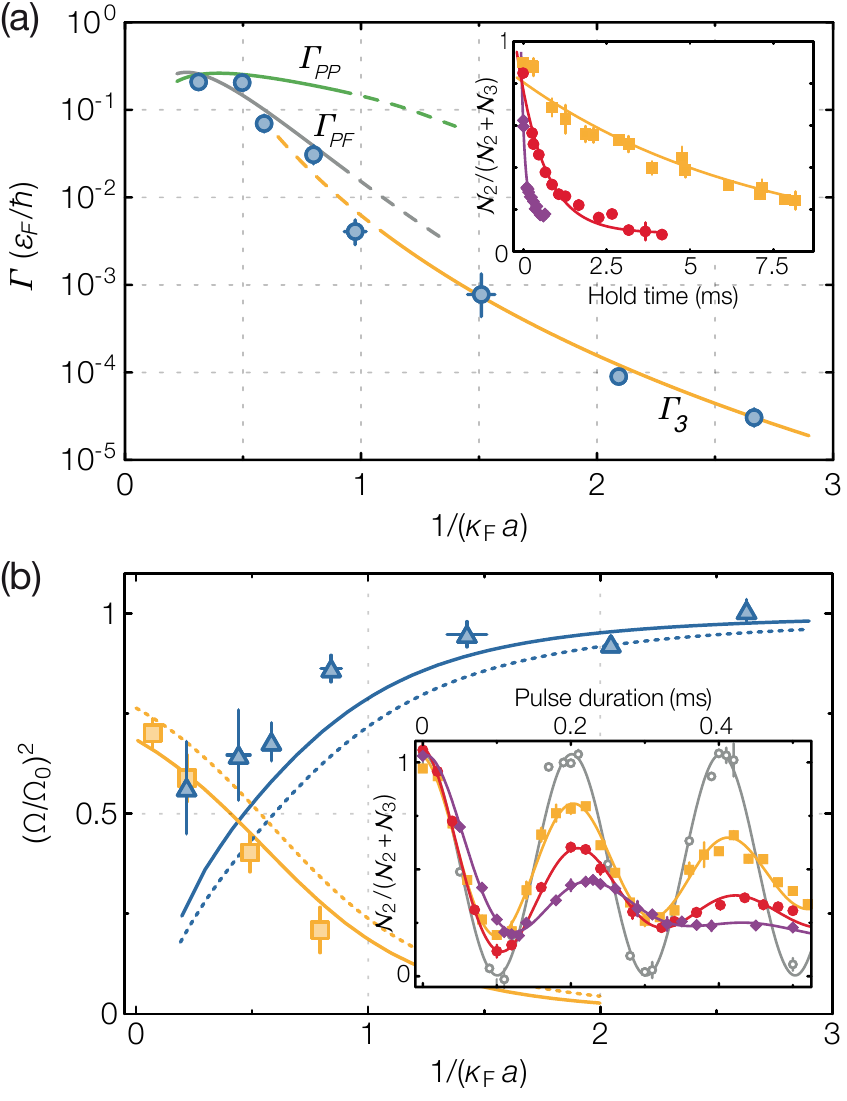}
\caption{
(a) Decay rate $\Gamma$ of the repulsive branch population measured as a function of $1/(\kappa_F a)$. Theory predictions for three-body recombination $\Gamma_3$ \cite{Petrov2003} (yellow line), polaron-to-polaron $\Gamma_{\text{PP}}$ \cite{Massignan2011} (green line) and polaron-to-bare atom $\Gamma_{\text{PF}}$ \cite{SM} (gray line) decay processes are plotted within their respective regimes of validity. Inset: examples of polaron population decay for $\kappa_F a \simeq$ 1 (yellow squares), 1.3 (red circles), 3 (purple diamonds), together with the exponential fits. (b) $(\Omega/\Omega_0)^2$ for the repulsive (blue triangles) and attractive (yellow squares) polarons at various $1/(\kappa_F a)$.  
Solid curves are our theory predictions for $(\Omega/\Omega_0)^2$ obtained within the ladder approximation \cite{SM}, while dotted curves depict the lowest order results $\sqrt{Z_{+ 2} \, Z_{\pm 3}}$.
Inset: repulsive polaron Rabi oscillations at $x=0.15(3)$ for $\kappa_F a \simeq $ 0 (empty grey circles), 1.1 (yellow squares), 1.3 (red circles), 1.7 (purple diamonds). Error bars combine the fit parameter errors with binned data s.e.m.}
\label{Fig4}
\end{center}
\end{figure}

After investigating the elastic properties of the repulsive polaron, we turn to consider its lifetime. This is a key quantity that sets the stability of the repulsive Fermi gas, and the applicability of Landau's quasiparticle theory in its description. Following Ref.~\cite{Kohstall2012}, we measure the quasiparticle decay rate through a double-pulse excitation scheme: a first $\pi$-pulse initially transfers state-\2 atoms to the repulsive branch; a second pulse, identical to the first, selectively brings repulsive quasiparticles back to the weakly-interacting state after a variable hold time.  
By fitting the relative state-\2 population measured as a function of time with an exponential decay, we extract the decay rate $\Gamma$ of the repulsive quasiparticle branch for various interaction strengths. The results are summarized in Fig.~\ref{Fig4}(a). 
We do not observe any appreciable dependence of $\Gamma$ upon concentration 
for $0.05 \leq x \leq 0.4$, and the data shown in Fig.~\ref{Fig4} are collected setting $x=0.15(1)$. $\Gamma$ is found to strongly increase towards unitarity, spanning nearly 4 orders of magnitude in the range $1/(\kappa_F a) = 0.3\dots2.7$. 

At weak coupling $\kappa_F a < 1$, the polaron decay is well described by the three-body recombination at rate $\Gamma_3$ of bare impurities colliding with two fermions from the bath \cite{Petrov2003}. For  $\kappa_F a > 1$, the medium starts playing an essential role: the bound-state energy is modified with respect to the one in vacuum, and alternative decay channels open up \cite{SM, Massignan2014}. In particular, it has been predicted that the zero-momentum repulsive polaron lifetime is limited for strong interactions by rapid two-body inelastic decay onto attractive polarons, which represent the many-body ground state for $\kappa_F a \geq 1.15$. The two-body decay rate $\Gamma_{\text{PP}}$, calculated in the ladder approximation for such polaron-to-polaron decay processes \cite{Massignan2011}, matches the data only close to unitarity, while greatly underestimating the polaron lifetime for $1/(\kappa_F a) > 0.5$. 
In contrast, we recover a good quantitative agreement for a wider range of $\kappa_F a$ by considering bare particles, rather than attractive polarons, as final decay products in the calculation \cite{SM} (see gray line in Fig.~\ref{Fig4}(a)).
In particular, we emphasize that the measured rate is about 20\% of $\varepsilon_F/\hbar$ close to unitarity, and it is below 10\% of $\varepsilon_F/\hbar$ at $1/(\kappa_F a)_c \simeq 0.6$, much smaller than theoretical expectations \cite{Massignan2011, Schmidt2011}. 

\enlargethispage{\baselineskip}
Finally, we probe the coherence properties of the repulsive polaron. As opposed to molecular excitations, polaron quasiparticles feature a coherent nature, usually quantified in terms of the quasiparticle residue $Z$ \cite{Massignan2014}, namely the squared overlap between the non-interacting and the many-body polaron wavefunctions. Information on $Z$ can be obtained either from spatially-resolved RF spectra \cite{Schirotzek2009} or by driving Rabi oscillations on the free-to-polaron transition \cite{Kohstall2012}. 
For a non-interacting initial state, $Z=(\Omega/\Omega_0)^2$, where $\Omega$ ($\Omega_0$) is the Rabi frequency of the polaron quasiparticle (bare particle) state \cite{Kohstall2012}. We modify this simple relation to account for the non-zero interactions of the  initial \2-state with the bath, and obtain predictions for $(\Omega/\Omega_0)^2$ \cite{SM}. 
Examples of repulsive polaron Rabi oscillations at various $1/(\kappa_F a)$ values are displayed in the inset of Fig.~\ref{Fig4}(b), where $\Omega_0 = 2\pi \times 4.95(5)\,\text{kHz} \simeq 0.7\,\varepsilon_F/\hbar$. As $\kappa_F a$ is increased, the repulsive polaron Rabi frequency progressively decreases, accompanied by an increasing damping rate of the oscillations. Such damping is much faster than the corresponding quasiparticle decay, indicating that decoherence induced by elastic collisions, rather than inelastic relaxation processes, is the dominant damping mechanism \cite{Kohstall2012, Cetina2016, SM}. Interestingly, the damping rate quantitatively matches the predicted quasiparticle peak spectral width \cite{Schmidt2011, SM}. 
The extracted $(\Omega/\Omega_0)^2$ from damped sinusoidal fits for both repulsive and attractive polarons are presented in Fig.~\ref{Fig4}(b), together with our theoretical predictions based on the ladder \mbox{approximation}~\cite{SM}. 

In conclusion, we presented a thorough study of the elastic and inelastic properties of repulsive Fermi polarons for a mass-balanced highly-polarized spin mixture at a broad Feshbach resonance.
While further theoretical effort is required for a comprehensive description of our experimental data, we demonstrate repulsive quasiparticle lifetimes greatly exceeding $10\,\hbar/\varepsilon_F$ over a wide range of interactions, far longer than recent predictions \cite{Massignan2011, Schmidt2011}. We also show that repulsive polarons remain well-defined coherent excitations even at very strong coupling by observing Rabi oscillations up to $1/(\kappa_F a) \simeq 0.2$. Moreover, we reveal an interaction regime where the paramagnetic Fermi liquid becomes energetically and thermodynamically unstable, motivating future studies aimed at directly observing ferromagnetism in metastable repulsive Fermi gases. Finally, our spectroscopic protocol can be extended to balanced mixtures, opening up new perspectives for monitoring the dynamical growth of polarized domains after a fast, yet selective RF quench to the upper branch of the many-body system, and the competing pairing instability \cite{Massignan2014, Pekker2011, Valtolina2016, Sanner2012}.

\begin{acknowledgments}
We thank G. Bertaina, G. Bruun, X. Cui, T. Enss, O. Goulko, D. Petrov, S. Pilati, N. Prokof'ev, B. Svistunov, H. Zhai, and the LENS Quantum Gases group for useful discussions, and R. Grimm  and J. Levinsen for a critical reading of the manuscript. This work was supported by the ERC through grants no.\:307032 QuFerm2D and no.\:637738 PoLiChroM, and through the EU H2020 Marie Sk\l{}odowska-Curie program (fellowship to F.S.).
P.M. acknowledges funding from a ``Ram\'on y Cajal" fellowship, from MINECO (Severo Ochoa SEV-2015-0522, and FOQUS FIS2013-46768), Generalitat de Catalunya (SGR 874), and the Fundaci\'o Privada Cellex.
\end{acknowledgments}

\smallskip
\textit{Note added.} -- While completing the experimental measurements, we became aware of related theoretical work by Goulko \textit{et al.}~\cite{Goulko2016}, in which diagrammatic Monte-Carlo results are shown to be compatible with our data.

\vspace*{35pt}
\bibliography{../Polarons}

\newpage
\newpage

\onecolumngrid
\begin{center}



\newpage\textbf{
Supplemental Material\\[4mm]
\large Repulsive Fermi polarons in a resonant mixture of ultracold $^6$Li atoms}\\
\vspace{4mm}
{F.~Scazza,$^{1,2,*}$ 
G.~Valtolina,$^{1,2}$
P.~Massignan,$^{3}$
A.~Recati,$^{4,5}$
A.~Amico,$^{2}$\\
A.~Burchianti,$^{1,2}$
C.~Fort,$^{2}$
M.~Inguscio,$^{1,2}$
M.~Zaccanti,$^{1,2}$
and G.~Roati$^{1,2}$}\\
\vspace{2mm}
{\em \small
$^1$Istituto Nazionale di Ottica del Consiglio Nazionale delle Ricerche (INO-CNR), 50019 Sesto Fiorentino, Italy\\
$^2$\mbox{LENS and Dipartimento di Fisica e Astronomia, Universit\`{a} di Firenze, 50019 Sesto Fiorentino, Italy}\\
$^3$ICFO-Institut de Ciencies Fotoniques, The Barcelona Institute of Science and Technology, 08860 Castelldefels, Spain\\
$^4$INO-CNR BEC Center and Dipartimento di Fisica, Universit\`a di Trento, 38123 Povo, Italy\\
$^5$Ludwig-Maximilians-Universit\"at M\"unchen, 80333 M\"unchen, Germany\\}
{\small$^*$ E-mail: scazza@lens.unifi.it}
\end{center}
\setcounter{equation}{0}
\setcounter{figure}{0}
\setcounter{table}{0}
\setcounter{section}{0}
\setcounter{page}{1}
\makeatletter
\renewcommand{\theequation}{S.\arabic{equation}}
\renewcommand{\thefigure}{S\arabic{figure}}
\renewcommand{\thetable}{S\arabic{table}}
\renewcommand{\thesection}{S.\arabic{section}}



\section{Experimental methods}

\subsection{Preparation of the sample}
By following procedures described in detail in Refs.~\cite{Burchianti2014, Valtolina2016}, we produce a weakly interacting, degenerate Fermi mixture of $^6$Li atoms held in a crossed optical dipole trap, at a temperature $T/T_F=0.10(2)$. The atoms composing the mixture symmetrically populate the lowest and third-to-lowest Zeeman states. These are characterized at low magnetic field by quantum numbers \ket{F=1/2, m_F=+1/2} and \ket{F=3/2, m_F=-3/2}, and they are denoted in the main text and in the following as \1 and \3 respectively. The state labeled as \2 corresponds instead to the hyperfine state \ket{F=1/2, m_F=-1/2}. 
Efficient evaporation is achieved by setting the bias magnetic field at 300\,G, where the interspecies scattering length is relatively large, $a_{13}\simeq -900\,a_0$, though non resonant \cite{Burchianti2014}. The final trap is cigar-shaped, with axial and radial frequencies $\omega_\text{ax} = 2\pi \times 19.7(2)$\,Hz and $\omega_\bot = 2\pi \times 233(5)$\,Hz, respectively.
To create a population-imbalanced $1-2$ mixture, we first adiabatically ramp the Feshbach field to 585\,G, where $a_{13} \simeq a_{12} \simeq +300\,a_0$. Then, by applying a radio-frequency (RF) pulse of adjustable duration, we transfer a certain fraction of state-\3 atoms into state \2. Immediately after, a 3\,$\mu$s-long optical blast selectively removes all remaining state-\3 atoms without causing appreciable heating on the remaining 1-2 sample. 
By varying the RF pulse duration we can precisely adjust the population $N_2$ of the minority component, correspondingly tuning the relative concentration $x= N_2/N_1$. For the various measurements presented in the main text, $x$ is varied between 0.05 and 0.4. At the end of this procedure, the majority and minority component gases are found in thermal equilibrium, irrespective of the specific value of $x$.
Once the desired population imbalance is obtained, we increase the bias field to values between 600 and 690\,G, a range where two off-centered 1-2 and 1-3 Feshbach resonances allow for resonantly tuning the interspecies scattering length $a_{13}$ on top of a weak increase of $a_{12}$ \cite{Zurn2013}. The final value of the magnetic field is accurately calibrated by driving the $\2 \rightarrow \3$ transition in a spin polarized gas containing only state-\2 atoms with a 1\,ms-long RF pulse, and determining its frequency with an uncertainty below 30\,Hz. The values of $a_{12}$ and $a_{13}$ at each bias field value are taken from Ref.~\cite{Zurn2013}, and are shown in Fig.~1a of the main text. 

\subsection{Reverse radio-frequency spectroscopy and Rabi oscillation measurements}\label{RabisSec}
In order to probe the spectral response of the strongly interacting mixture we employ reverse radio-frequency (RF) spectroscopy \cite{Gupta2003, Regal2003a, Kohstall2012}, applying RF pulses near-resonant with the $\2 \rightarrow \3$ transition, in the presence of the medium of state-\1 majority atoms. The minority component is thus transferred from a weakly interacting regime into a resonantly interacting state, allowing to probe not only the energy of the interacting ground state, but also of all higher-lying excitations including repulsive polarons. Our spectroscopic signal, as a function of the RF detuning $\Delta$ (measured with respect to the frequency of the bare $\2 \rightarrow \3$ transition), is defined as the ratio between the transferred atoms $\mathcal{N}_3$ and the total population of minority atoms $\mathcal{N}_2 +\mathcal{N}_3$, contained within the central region shown in the inset of Fig.~1a in the main text.
The populations $\mathcal{N}_2$ and $\mathcal{N}_3$ are monitored through absorption imaging, by shining two consecutive imaging pulses selectively resonant with the $\2$ and $\3$ atoms at the final field, separated by 500\,$\mu$s. 

For the repulsive polaron spectroscopy presented in the main text, we employ rectangular RF pulses, whose duration and power are adjusted to correspond to a 0.8\,$\pi$-pulse for non-interacting atoms. For most of the measurements, except those acquired in the strongly interacting regime $\kappa_F a_{13}>2$, the pulse length is set to 1\,ms, corresponding to a spectral width $\sim 0.13\,\varepsilon_F$. For the spectra recorded at $\kappa_F a_{13}>2$, the duration is reduced to 0.5\,ms while maintaining the 0.8\,$\pi$-pulse condition by increasing the RF power. While this in principle reduces the spectral resolution, it allows to increase the peak signal and limits undesired dynamical effects during the spectroscopy. In particular, a short spectroscopic pulse-duration suppresses appreciable contributions from 2-3 interactions: these can arise only if superpositions of states \ket{2} and \ket{3} are driven by impurity-bath collisions into statistical mixtures after a sufficient decoherence time.  

For the measurement of polaron Rabi oscillations, we set the RF frequency at the previously determined $\Delta_+$ ($\Delta_-$) of the repulsive (attractive) polaron. The RF power is set to the maximum value allowed by our apparatus, yielding a $\pi$-pulse duration of 100\,$\mu$s for non-interacting state-\2 atoms in the field region between 500 and 700\,G. The analysis of repulsive polaron Rabi oscillation data, that allowed to extract the frequency $\Omega/\Omega_0$ and the damping rate $\gamma_R$ is based on the fitting function $f(t) = A\,e^{-\Gamma_R t} + B\,e^{-\gamma_R t} \cos(\Omega t)$, describing an oscillation at frequency $\Omega/(2\pi)$ with a damping $\gamma_R$ and an excited-state population decay rate $\Gamma_R$ (with $A, B \simeq 0.5$) (see also Section~\ref{SecDecoh}). On the other hand, an excellent fit to the attractive polaron data is obtained by setting $\Gamma_R=0$.

\subsection{Measurement of repulsive polaron decay}\label{DecaySec}
The measurement of the decay rate of the repulsive polaron population is based on a double-pulse protocol, analogous to the one successfully exploited for the study of ${}^{40}$K impurities in a \Li Fermi sea \cite{Kohstall2012}. The measurement sequence starts by initially transferring impurity \2 atoms to state \3 by means of a fast RF $\pi$-pulse lasting between 100 and 200\,$\mu$s, whose frequency is adjusted to the previously determined repulsive polaron resonance, and whose power and duration are optimized to lead to the maximum transfer efficiency. After some variable hold time, a second RF pulse identical to the first selectively transfers back into state $\2$ only those impurities that still occupied the repulsive polaron branch. By recording the fraction of atoms which is transferred back to state \2 as a function of time, we directly probe the time evolution of the population of the upper branch, which for all regimes investigated was found to exhibit an exponentially decaying trend. Imaging of the $\3$ atoms provides in turn information on the population of molecules, or attractive polarons, onto which the population of the repulsive branch has decayed. We note here that our data analysis assumes a bath average density which does not vary over time. This is not strictly true, since the impurity concentration is small but finite. However, this would eventually lead to an overestimation of the real decay rate, therefore not changing the conclusions discussed in the main text.

\section{Effective Fermi energy and Fermi wavevector}\label{effEF}

In order to reduce the effects of density inhomogeneity of trapped samples on the spectroscopy signal, we only record the transferred atom fraction from the column density within a central rectangular region of size 30\,$\mu$m $\times$ 70\,$\mu$m along the transverse and axial directions of the cloud, respectively.
The majority Fermi gas, i.e. the bath, is characterized within this integration region by an effective Fermi energy $\varepsilon_F$ and an effective Fermi wavevector $\kappa_F$. 
$\varepsilon_F$ and $\kappa_F$ represent the mean value of the local $k_F(\br,T/T_F) = (6 \pi^2 n(\br,T/T_F))^{1/3}$ and $E_F(\br,T/T_F) = \hbar^2/(2m)\,k_F^2(\br,T/T_F)$, respectively, averaged over the integration region. These quantities provide the relevant length and energy scales, used for presenting our experimental findings and for comparing them with theoretical predictions obtained for a homogenous bath. Owing to the finite temperature and the residual trap inhomogeneity, the density of the bath in the integration region differs from the peak density of a zero-temperature Fermi gas, and therefore $\varepsilon_F$ and $\kappa_F$ significantly differ from the zero-temperature central Fermi energy $E_F$ and Fermi wavevector $k_F$. We can write $\varepsilon_F=\eta\,E_F$, where $\eta$ is a factor that depends on the region size and on the cloud temperature, and for the sufficiently small chosen region we have $\kappa_F \simeq \eta^{1/2}\,k_F$.

\begin{figure}[t!]
\begin{center}
\includegraphics[width= 8.6 cm]{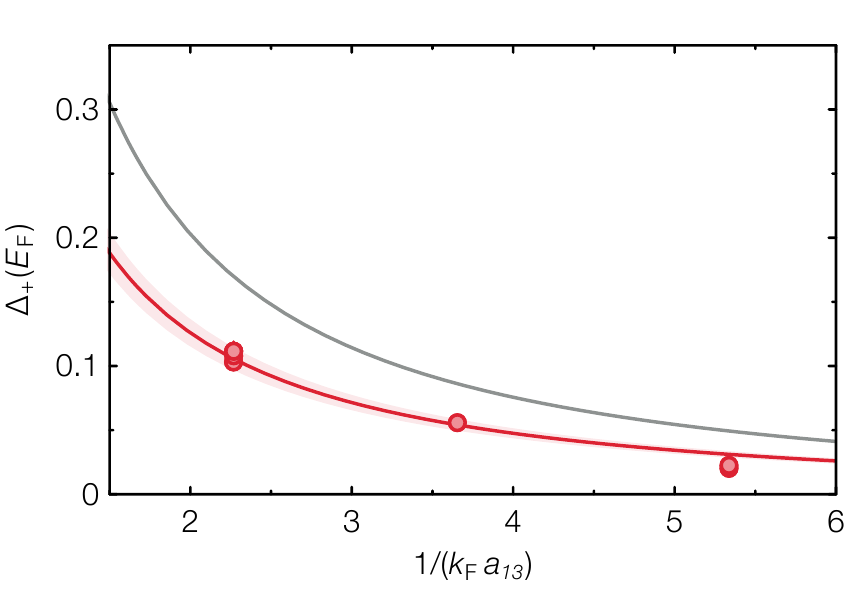}
\caption{Calibration of the effective Fermi energy $\varepsilon_F$ and effective Fermi wavevector $\kappa_F$. The measured RF shifts (red circles) are fitted with Eq.~\eqref{RFshiftLHY} (solid red line) to yield $\eta=0.74(2)$. The shaded red area denotes the standard confidence band of the fit. The curve obtained by setting $\eta=1$ in Eq.~\eqref{RFshiftLHY} is shown for comparison (solid gray line).}
\label{FigS1}
\end{center}
\end{figure}

In order to calibrate $\varepsilon_F$ and $\kappa_F$, we exploit the RF spectroscopy data at weak repulsion, where the RF shift $\Delta_+$ is given with high accuracy by:
\begin{equation}
\Delta_{+, \text{LHY}}= \varepsilon_F\left (\frac{4}{3 \pi} \,\kappa_F \,(a_{13}-a_{12})+ \frac{2}{\pi^2} \,\kappa_F^2 \,(a_{13}^2-a_{12}^2)\right)
\label{RFshiftLHY}	
\end{equation}
This result is obtained by treating perturbatively the impurity-bath interactions and including second-order beyond mean-field corrections \cite{Bishop1973}, namely the interaction energy is expanded up to second order in $\kappa_F a$. For $\kappa_F a \lesssim 0.5$, the second-order perturbative formula well approximates the results of QMC calculations \cite{Pilati2010} and variational or T-matrix methods \cite{Massignan2011, Cui2010}.

We can extract $\varepsilon_F$ and $\kappa_F$ by fitting the observed RF shifts at weak coupling $k_F a_{13} \leq 0.45$ with Eq.~\eqref{RFshiftLHY} (see Fig.~\ref{FigS1}), with a single fitting parameter $\eta$. From this analysis, we obtain $\eta=0.74(2)$.
Note that such a calibration of $\varepsilon_F$ and $\kappa_F$ does not depend on the knowledge of $E_F$ and $k_F$, since the measured RF shifts are directly compared to Eq.~\eqref{RFshiftLHY}. Nonetheless, the value of $\eta$ provides a factor that can be used to scale the spectroscopy shifts from experimental runs with slightly differing atom numbers (by up to 15\%).

The determined value of $\eta$ is compatible with another estimate $\eta_{int}=0.72(1)$, which we obtain by numerically computing the density-weighted average Fermi energy in the integration region in the local density approximation (LDA). We approximate the finite-temperature density distribution $n(\br)$ of the bath as the one of an ideal Fermi gas, imposing the measured $T/T_F$, trap frequencies and atom number $N$, i.e.:
\begin{align}
n(\br) &= - \left(\frac{m k_B T}{2\pi \hbar^2}\right)^{3/2} \text{Li}_{\frac{3}{2}}\left(-e^{\,\beta (\mu(T/T_F, N) - U(\br))}\right)\:.
\label{localEF}
\end{align}
Here, $\mu(T/T_F, N)$ is the central chemical potential, $U(\br)$ is the harmonic trapping potential, $\beta = 1/(k_B T)$, and Li${}_{s}$ stands for the polylogarithm function of order $s$.
$\varepsilon_F$ is obtained by averaging over the density profile within the integration region, denoted as $V$:
\begin{align}
\varepsilon_F &= \frac{\hbar^2}{2m\,\mathcal{N}} \int_V \text{d}\br \left ( 6\pi^2 n(\br) \right )^{2/3} n(\br)\:,
\end{align}
where $\mathcal{N} =  \int_V \text{d}\br n(\br)$ is the number of majority atoms in the region. \\
Since the latter estimate is intrinsically affected by the uncertainty in determining the cloud temperature and neglects the effect of weak 1-2 repulsive interactions on the density distribution of the bath, we adopt the value of $\eta$ directly obtained from fitting the RF shifts at weak interactions with Eq.~\eqref{RFshiftLHY}, which only relies on the precise knowledge of the scattering lengths $a_{12}$ and $a_{13}$ \cite{Zurn2013}. Yet, through the density-profile integration we estimate the standard deviation of the local Fermi energy in the chosen integration region $\Delta \varepsilon_F \simeq 0.2\,\varepsilon_F$ and the the standard deviation of the local Fermi wavevector $\Delta \kappa_F \simeq 0.15\,\kappa_F$. 


\section{Mean energy per particle of the impurity gas}\label{SecMeanEpsilon}
\begin{figure}[h!]
\begin{center}
\includegraphics[width= 8.6 cm]{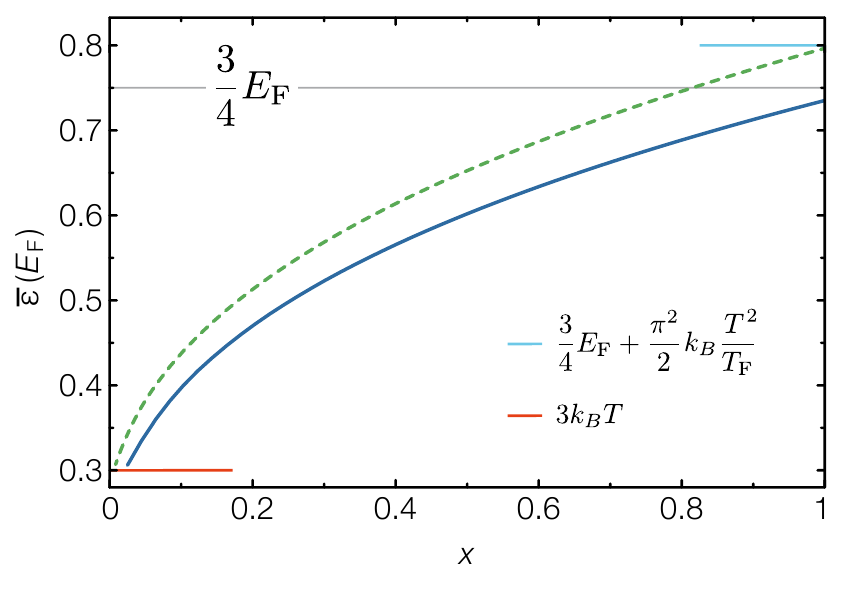}
\caption{Mean energy per particle in the experimental spectroscopy integration region $V$. The solid blue line is the result of Eq.~\eqref{EReg} for $T/T_F = 0.1$. For comparison, the result of Eq.~\eqref{EReg} when $V$ is set to the full trap volume is plotted as a dashed green line. 
}
\label{FigS2}
\end{center}
\end{figure}

In order to extract the polaron energy and effective mass by a linear fit of the data with Eq.~(1) of the main text, the relevant effective quantity is the mean energy per impurity $\bar{\varepsilon}$ rather than the impurity concentration $x$. By increasing  $x$, $\bar{\varepsilon}$ increases owing to the modification of the impurities Fermi pressure. We thus need to evaluate $\bar{\varepsilon}$ as a function of $x$ for our integration region $V$.
For this purpose, we compute the average energy of the minority state-\2 gas in the integration region $V$, assuming state-\2 is non-interacting, but imposing thermal equilibrium with the state-\1 bath, i.e. $T_2 = T \simeq 0.1\,T_F$. 
In the local density approximation (LDA), the mean energy per particle as a function of $x$ is obtained as:
\begin{equation}
\bar{\varepsilon} (x) \equiv \bar{\varepsilon}_2 (x) = \frac{4\pi}{(2\pi \hbar)^3\,\mathcal{N}_2(x)} \int_V \text{d}\br \int_0^{\infty} dp\,p^2\, \frac{\varepsilon_{\bp,\br}}{e^{ \beta \left (\varepsilon_{\bp,\br} - \mu(x) \right)} + 1}\:,
\label{EReg} 	
\end{equation}
where $\mathcal{N}_2(x) =  \int_V \text{d}\br \,n(\br; x)$ is the number of impurity atoms in the integration region. Here, $\varepsilon_{\bp,\br} = p^2/2m + U(\br)$ is the single-particle energy and $\mu(x)$ is the chemical potential of the impurities. The degree of degeneracy of the impurities decreases as $x$ decreases from 1 to 0, and consequently $\mu(x)$ decreases.
The first integration over momentum can be performed analytically, yielding:
\begin{equation}
\bar{\varepsilon} (x) = - \frac{1}{\mathcal{N}_2(x)}\left(\frac{m k_B T}{2\pi \hbar^2}\right)^{3/2} \int_V \text{d}\br  \left(\frac{3}{2}\,k_B T\, \text{Li}_{\frac{5}{2}} \left(-e^{\beta  (\mu(x) -U(\br))}\right)+U(\br)\, \text{Li}_{\frac{3}{2}}\left(-e^{\beta (\mu(x) -U(\br))}\right)\right)\:.
\end{equation} 
The integration over $V$ is then performed numerically, yielding as a function of $x$ the curve plotted as a solid line in Fig.~\ref{FigS2}. 

\section{Theoretical methods}
\subsection{Polaron properties}

A single spin-down atom (the $\down$ ``impurity") in an ideal Fermi gas of spin-up atoms forms a quasi-particle, commonly known as the ``Fermi polaron" \cite{Chevy2006, Prokofev2008, Schirotzek2009, Massignan2014}. 
The microscopic interaction potential between ultracold atoms is always attractive, leading to a negative energy for the ground-state polaron.
Beyond a critical attraction, the ground state of the mixture becomes a dressed molecule. Associated with this latter quasiparticle, a broad continuum appears in the impurity spectral function, since the impurity can bind to any fermion of the bath with kinetic energy ranging between 0 and $E_F$.
Moreover all the approaches to the problem and various experiments revealed that an additional quasi-particle excitation, termed ``repulsive polaron'', appears at positive energies, well separated from the attractive polaron even in the strongly interacting regime \cite{Cui2010, Massignan2011, Schmidt2011, Kohstall2012, Koschorreck2012}.

The Green's function for the impurity reads (we set in the following of this Section $\hbar=k_B=1$)
\beq
G_\down(\bp,\omega)={1 \over \omega-\xi_{\bp\down}-\Sigma(\bp,\omega)+i0^+},
\eeq
where $\xi_{\bp\sigma}=\varepsilon_{\bp\sigma}-\mu_\sigma=p^2/2m_\sigma-\mu_\sigma$ is the kinetic energy of a $\sigma$ atom measured with respect to its chemical potential. The effects of the interactions with the Fermi sea are contained in the retarded self-energy $\Sigma(\bp,\omega)$.\\
For well defined polarons one can expand the Green's function around the real part $E_\pm$ of the pole at $p=0$, where $+$ ($-$) refers to repulsive (attractive) polarons.  The approximated Green's function at momenta $p\ll \kappa_F$ reads
\beq\label{eq:polaronGreensFunction}
G_\down(\bp,\omega)
\approx\frac{Z_\pm}{\omega-E_\pm-\frac{p^2}{2m^*_\pm}+i\Gamma_\pm/2},
\eeq
where the quasiparticle residues $Z_\pm$ are defined as
\beq
Z_\pm=\frac{1}{1-\mathrm{Re}[\partial_\omega\Sigma(0,E_{\down\pm})]},
\eeq
the effective masses are given by
\beq
m^*_\pm= {m_\downarrow/Z_\pm \over 1+\mathrm{Re}[\partial_{\varepsilon_{\down\bp}}\Sigma(0,E_{\downarrow\pm})]},
\eeq
and the decay rate of the quasiparticle's probability density is given by
\beq
\Gamma_\pm=-2 Z_\pm{\rm Im}[\Sigma(\bp,E_{\downarrow\pm})].
\eeq
While the attractive polaron, being the ground state of the many-body system until the transition to the molecular state, is a stable quasiparticle, the repulsive polaron is always metastable, but it remains a well-defined quasi-particle as long as $\Gamma_+/2$ is much smaller than $E_+$.

\smallskip
In the spirit of Landau Fermi liquid theory, the change in the quasiparticle energy associated with a finite though low density of impurities, $x = N_\downarrow/N_\uparrow \ll 1$, can be accounted for by writing an energy functional of the many-body system with two concentration-dependent terms. For a zero-temperature homogeneous mixture in thermodynamical equilibrium this reads \cite{Lobo2006, Pilati2008}:
\beq\label{energyFunctional}
\frac{E-\frac{3}{5}E_F N_\up}{N_\down}= E_\pm + \frac{3}{5}E_F \frac{m}{m^*_\pm}x^{2/3}+\frac{3}{5}E_F F_\pm x\:.
\eeq 
where $E_F=k_{F\up}^2/2m_\up$ is the Fermi energy of the majority component. The term $\propto 1/m^*$ accounts for the Fermi pressure of the quasiparticle Fermi gas, while the latter term proportional to the Landau parameter $F$ can be viewed as an effective impurity-impurity interaction. The Landau parameter $F$ can be expressed as \cite{Mora2010, Yu2010, Yu2012a}
\beq
F_{\pm}=\frac{5}{9}\left(\Delta N_{\pm}\right)^2\:,
\eeq
where $\Delta N_{\pm}=\partial n_{\down,\pm}/\partial n_\up \approx-\partial E_{\pm}/\partial E_F$ is the number of majority particles in the dressing cloud of a polaron \cite{Massignan2011}. Note that, by its own definition, the Landau parameter $F$ is always positive. Namely, effective polaron-polaron interactions are always repulsive.

\begin{figure}[b!]
\centering
\includegraphics[width= 12 cm]{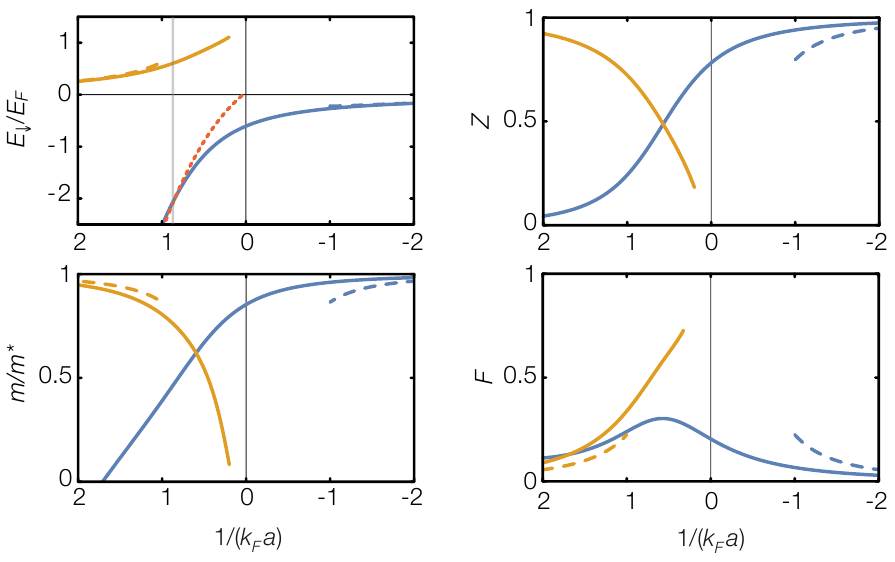}
\caption{\label{fig:qp_properties}
Polaron properties calculated in the ladder approximation for the mass-balanced case: energy $E_\down$, residue $Z$, inverse effective mass $m/m^*$ and Landau parameter $F$. Blue and orange lines depict, respectively, the attractive and the repulsive polaron.
The dotted red line is the energy of a dressed molecule, and the thin vertical line at $k_Fa\simeq1.15$ indicates the polaron/molecule transition in the attractive branch \cite{Prokofev2008, Vlietinck2013}.
The dashed lines are the perturbative results to $O(k_Fa)^2$, as given in Refs.~\cite{Bishop1973, Trefzger2013, Mora2010}.
}
\end{figure}

\smallskip
The trend associated with the energy functional Eq.\ \eqref{energyFunctional} can be qualitatively captured if one considers repulsive polarons in the weak-coupling limit, where Eq.\ \eqref{energyFunctional} can be expanded in powers of the small parameter $\kappa_F a$. At second order it reads\\[2mm]
\beq\label{energyFunctional2ndOrder}
\frac{E-\frac{3}{5}E_F N_\up}{N_\down E_F}= \left(\frac{4}{3\pi} k_Fa+\frac{2}{\pi^2}( k_Fa)^2\right)
+ \frac{3}{5} \left(1-\frac{4}{3\pi^2}(k_Fa)^2\right)x^{2/3}
+\frac{3}{5} \left(\frac{20}{9\pi^2}(k_Fa)^2\right) x +O(k_Fa)^3\:,\\[2mm]
\eeq
The kinetic energy term $\propto x^{2/3}$ is systematically smaller than the corresponding one associated with bare particles, and it therefore causes a decrease of the RF shifts for increasing $x$, with a trend fully consistent with the one observed for the repulsive polaron in the experiment. An opposite behaviour is given by the effective interaction term, which is always positive for the quasiparticles and zero for the bare ones: this would lead to a linear increase of the quasiparticle energy with increasing $x$, clearly in contrast to what is measured.
It is important to emphasize that the competition between kinetic and effective interaction terms in Eq.~\eqref{energyFunctional2ndOrder} would lead to a non-monotonic trend of the RF shift, featuring a decrease for small $x$ followed by an increase once the interaction term becomes dominant. This qualitative trend holds also at strong repulsion, where the Landau parameter $F$ can be computed within the ladder approximation (see Fig.\ \ref{fig:qp_properties}). Despite the finite temperature and density inhomogeneity of our system could quantitatively affect the aforementioned behavior, effective polaron interactions would cause the RF shift to be neither monotonic in $x$ nor in $\epsilon$. Furthermore, if sizeable interaction effects appeared in our study, these should also affect the quasiparticle coherence properties; we never observe such effects in the experiment (see the concluding Sec.\ \ref{SecDecoh}).

\smallskip
We conclude that our setup does not show any effect of impurity-impurity interactions. We argue that this is due to the explored concentrations, which are sufficiently low to make polaron interactions not relevant (see Eq. (\ref{eq:F-RF}) and comment below it). Furthermore, in reverse RF spectroscopy one may expect that polaron-polaron interactions cannot be established over the limited timescale of the RF pulse. This hypothesis could share analogies with the recent studies about the different timescales to establish few-body correlations in a thermal Bose gas after an interaction quench \cite{Fletcher2016}: three-body correlations are found to grow over timescales greatly exceeding the ones associated with the two-body correlations.

\newpage
\bigskip
\noindent\textbf{Ladder approximation and one-particle-hole Ansatz}\\[2mm]
The calculation of the self-energy in presence of very strong interactions is in principle a formidable task. However it turns out that a very good approximation is obtained by summing only the so-called ``ladder diagrams", describing forward-scattering in the medium \cite{Combescot2007}. Under this approximation, the retarded self-energy of a single impurity in a zero-temperature Fermi sea close to a broad Feshbach resonance reads ($\hbar=k_B=1$)
\vspace*{2mm}
\begin{equation}\label{self_energy}
\Sigma(\mathbf{p},\omega)=\sum_{\bq} f(\xi_{\bq\up})\,T(\bp+\bq,\omega+\xi_{\bq\up})
=\sum_{\bq}\frac{f(\xi_{\bq\up})}
{\frac{m_r}{2\pi a}-\sum_{\bk}\left[\frac{1-f(\xi_{\bk\up})}{\omega-(\varepsilon_{\mathbf{p+q-k}\down}+\varepsilon_{k\up}-\varepsilon_{q\up})+i0_+}+\frac{2m_r}{k^{2}}\right]},
\end{equation}\\[0.3mm]
where $f(x)=1/(\exp(x/T)+1)$ is the Fermi function, and $T(\mathbf{P},\Omega)$ is the T-matrix describing the scattering of an $\up-\down$ pair of atoms with total momentum $\mathbf{P}$ and total energy $\Omega$. Furthermore, we have introduced the $\up-\down$ scattering length $a$, and the reduced mass  $m_r=m_{\uparrow}m_{\downarrow}/(m_{\uparrow}+m_{\downarrow})$. 
The quasiparticle properties obtained for the attractive polaron through this ladder approximation compare very favorably both with Quantum Monte-Carlo (QMC) calculations, and with experiments \cite{Prokofev2008, Schirotzek2009, Kohstall2012, Massignan2014}. 

As the ground state properties of the system are considered, solving the problem within the ladder (or non self-consistent T-matrix) approximation is equivalent to minimizing the energy over  the  ``one-particle-hole" (1PH)  variational Ansatz \cite{Chevy2006, Combescot2007}:
\begin{equation}\label{1PH_ansatz}
|\psi\rangle= \sqrt{Z} |{\bf p}\rangle_\downarrow\;|0\rangle_\uparrow + \!
\sum_{q<\kappa_F}^{k>\kappa_F}\phi_{\bk \bq} |{\bf p}+{\bf q}-{\bf
  k}\rangle_\downarrow\;c^{\dag}_{{\bf k}\uparrow}\,c_{{\bf q}\uparrow}\,|0\rangle_\uparrow\:.
\end{equation}
Eq.~\eqref{1PH_ansatz} describes a $\downarrow$-impurity with momentum $\bp$ in an ideal Fermi sea $|0\rangle_\uparrow$ of $\uparrow$-particles as a quasiparticle, whose dressing is composed by a superposition of particle-hole pairs.

\enlargethispage{\baselineskip}
Since it is analytically continued, the ladder calculation has however the additional advantage of being able to investigate excited states of the mixture, such as the repulsive polarons discussed in this work. For the repulsive polaron case, such an approach is found in relatively good agreement with QMC simulations \cite{Pilati2010, Massignan2014}, despite a more sizeable mismatch is present with respect to the attractive polaron case. Nonetheless, the ladder approximation accurately reproduced the experimental results obtained in a \Li-${}^{40}$K mixture at a relatively narrow Feshbach resonance \cite{Kohstall2012}.
The basic properties of the attractive and repulsive polarons, such as their energy, quasiparticle residue, effective mass, and Landau parameter $F$ obtained by means of the ladder approximation, are summarized in Fig.\ \ref{fig:qp_properties}. 

The repulsive polaron energy from Ref.~\cite{Cui2010} depicted as a dot-dashed yellow line in Fig.~3a of the main text was derived by a variational treatment based on the 1PH Ansatz, Eq.~\eqref{1PH_ansatz}.
This procedure leads to an equation identical to Eq.~\eqref{self_energy}, except that only the principal part of the denominator of the integrand is retained (i.e., its imaginary part is neglected). This approach yields a purely real energy for the repulsive polaron, which equals $1.82\,E_F$ at resonance.

\bigskip
\noindent\textbf{Decay rates}\\[2mm]
The intermediate states contained in the ladder self-energy given in Eq.~\eqref{self_energy} represent bare particles, not dressed by interactions, and therefore the latter does not adequately describe the decay processes. 
On the other hand, it is possible to argue that the lifetime of a repulsive polaron is limited by its possibility of scattering with fermions of the bath, releasing its energy and decaying into an attractive polaron via a two-body process or into a molecule via a three-body process on the BEC side of the resonance. Such an approach has already provided accurate predictions for narrow resonances and heavy impurities \cite{Kohstall2012, Massignan2014}. In this case, the estimate for the two-body polaron-to-polaron decay rate (to which we refer in the main text as $\Gamma_{\rm PP}$) is based on the assumption that the bare matrix element for the decay of the repulsive polaron coincides with the two-body T-matrix in vacuum, taking into account the multiple scattering of the attractive polaron with the particle of the bath considering only the ladder diagrams. The resulting ``dressed" self-energy reads exactly as Eq.~\eqref{self_energy} but with the properties (mass, residue, energy) of the attractive polaron in the pair propagator.
 
In our broad resonance and homonuclear case, however, the energy difference between the repulsive and attractive polaron is generally larger than $E_F$, and the bare impurities have the same mass as the majority particles. Therefore the scattered impurities (i.e., the final states of the decay process) usually carry a high momentum $k\gtrsim \kappa_F$. As a consequence, their description in terms of the low-momentum quasiparticle properties given by Eq.~\eqref{eq:polaronGreensFunction} is rather poor, since it strongly overestimates the medium effects on the particle dressing. 
In particular, the effects of the bath eventually become negligible for very large momenta.
A better estimate can be thus obtained by assuming that the repulsive polaron decays into a bare impurity with a weight $1-Z_+$. Approximating $m^*_+\approx m$, the corresponding self-energy reads
\beq\label{dressed_self_energy}
\tilde\Sigma(\mathbf{p},\omega)
=\sum_{\bq}\frac{f(\xi_{\bq\up})}
{\frac{m}{4\pi a} - (1-Z_+)\sum_{\bk}\left[\frac{1-f(\xi_{\bk\up})}{\omega-(\varepsilon_{\mathbf{p+q-k}\down}+\varepsilon_{k\up}-\varepsilon_{q\up})+i0_+}+\frac{m}{k^{2}}\right]}\:,
\eeq 
and the polaron-to-bare particle population decay rate is given by
\beq\label{Gamma_PP}
\Gamma_{\rm PF}\equiv -2Z_+ {\rm Im}[\tilde\Sigma(0,E_+)]\:.
\eeq

A competing decay channel, when the molecule becomes the ground state of the impurity problem, is the three-body recombination of impurities with fermions of the bath ($\{\uparrow, \uparrow, \downarrow\} \rightarrow \{ \uparrow, \uparrow\downarrow \}$).
Deep in the BEC regime this process is expected to be quantitatively described by the three-body recombination rate determined by Petrov \cite{Petrov2003}:
\beq\label{GammaPetrov}
\Gamma_{3} = \frac{148\,\hbar\,a^4}{m} \frac{\bar{\epsilon}}{\epsilon_B}\,n_{\uparrow}^2\:.
\eeq
Here, $n_{\uparrow}$ is the density of the bath, $\epsilon_B$ is the molecule binding energy, and $\bar{\epsilon}$ is the mean collision energy, which in turn corresponds to the mean kinetic energy of a fermion of the bath. We neglect the collision energy contribution from the impurity, which is small when $x \ll 1$. 
We set $\epsilon_B \approx \hbar^2/(m a^2) + \varepsilon_F$, which accounts for the removal of one fermion from the bath once a molecule is formed, while it neglects atom-dimer interactions and non-trivial medium effects. From this we obtain:
\beq\label{GammaPetrovFiniteT}
\frac{\hbar\Gamma_{3}}{\varepsilon_F} = 0.298\,\frac{148}{(6 \pi^2)^2} \frac{(\kappa_F a)^6}{1+\frac{1}{2}(\kappa_F a)^2}\:,
\eeq
where the numerical factor $0.298$ results from averaging the kinetic energy and the squared density of the bath over the integration region. 
The results for $\Gamma_{\rm PP}$, $\Gamma_{\rm PF}$ and $\Gamma_3$ are compared to the experimentally measured decay rates in Fig.\ 4a of the main text.

\subsection{Reverse radio-frequency spectroscopy}

In order to extract the energy and the effective mass of the polaron we rely on the so-called reverse radio-frequency (RF) spectroscopy. The minority gas is prepared in state $|2\rangle$, weakly interacting with the majority component. We then apply a RF pulse to transfer the atoms to the empty state $|3\rangle$, which is resonantly interacting with the atoms of the bath. For a homogeneous system and within linear response theory, the RF signal is given by (see e.g. Ref. \cite{Massignan2008})
\beq
\label{RFhom}
I(\omega)\propto\int\frac{{\rm d}\bk}{(2\pi)^3}
f(\xi_{k,2})\, A_3(\bk,\xi_{k,3}+\omega),
\eeq
where $A_3(\bk,\omega)=-2\,{\rm Im}G_3(\bk,\omega)$ is the spectral function for the minority gas in state $|3\rangle$.  
Since the RF pulse has a finite duration, the previous result has to be convoluted in frequency as 
\beq
I_{exp}(\omega)=\int {d\varepsilon\over 2\pi} \, g(\omega-\varepsilon)\,I(\varepsilon),
\eeq
where $g(\omega)$ can be well approximated by a Gaussian whose width is inversely proportional to the duration of the RF pulse.

When the spectrum contains a well-defined quasiparticle with zero-momentum energy $E_\pm$, the main contribution to the spectral function comes from the quasiparticle pole, i.e., $A_3(\bk,\varepsilon)=2\pi Z\,\delta(\varepsilon-E_\pm-k^2/2m^*)$, and the RF signal reads
\beq
I_{exp}(\omega)\propto\int\frac{{\rm d}\bk}{(2\pi)^3}
f(\xi_{k,2})\,g(\Delta_k-\omega),\;\;\textrm{with }\Delta_k=E_\pm+{k^2\over 2m}\left({m\over m_\pm^*}-1\right).
\eeq
Therefore the first moment $\bar\omega$ of the RF signal is given by the average over the Fermi distribution of $\Delta_k$, i.e.
 \beq
\bar \omega=\frac{\int d\omega\; \omega\,I_{exp}(\omega)}{\int d\omega\; I_{exp}(\omega)}\propto
\frac{\int{\rm d}\bk\;\Delta_kf(\xi_{k,2})}{\int{\rm d}\bk\; f(\xi_{k,2})}=E_\pm+\langle{k^2\over 2m}\rangle\left({m\over m_\pm^*}-1\right).
\eeq
In our experiment, the signal $I_{exp}(\omega)$ is essentially Gaussian, so that its peak value coincides with $\bar\omega$.

Our system is however not homogeneous, since the atoms feel a harmonic trapping potential $U(\br)=m(\omega_x x^2+\omega_y y^2+\omega_z z^2)/2$. If $\mu_i$ is the chemical potential of the impurity atoms in state $|i\rangle$ the RF peak is located at $\mu_3-\mu_2$. Within LDA at $T=0$ we can write
\begin{eqnarray}
\mu_2&=&\frac{(6\pi^2\,n_2(\br))^{2/3}}{2m}+U(\br)\:,\\
\mu_3&=&\frac{(6\pi^2\,n_3(\br))^{2/3}}{2m_\pm^*(\br)}+E_\pm(\br)+U(\br)\:.
\end{eqnarray}
To estimate the polaron parameter we average the difference  $\mu_3-\mu_2$ on the non-interacting density $n_2$ on a finite region $V$. Since the variance of the majority Fermi momentum over the integration region $V$ is only around 1\% of its mean value $\kappa_F$ (see Section~\ref{effEF}), we may safely approximate the average values of the polaron parameters by their values computed at $\kappa_F$. Assuming furthermore that in such a region $n_3$ does not differ too much from $n_2$, we obtain
\beq\label{RFShift3}
\Delta_\pm=\langle \mu_3-\mu_2\rangle=
\frac{1}{\mathcal N_2}\int_V{\rm d}\br\left(\frac{(6\pi^2 n_3(\br))^{2/3}}{2m_\pm^*(\br)}-\frac{(6\pi^2 n_2(\br))^{2/3}}{2m}+E_\pm(\br)\right)n_2(\br)
\simeq E_\pm(\kappa_F a)+\bar \varepsilon\left({m\over m_\pm^*(\kappa_F a)}-1\right), \rule[-1.5em]{0pt}{0pt}
\eeq
\\[0.5mm]
with $\mathcal N_2=\int_V{\rm d}\br\; n_2(\br)$. The previous expression corresponds to Eq.~(1) given in the main text, where $\bar \varepsilon$ is the average energy of the state-\2 component in the region of integration $V$. At finite temperature the same expression holds, just using the proper equation of state for the state-\2 gas (see Section~\ref{SecMeanEpsilon}).

With similar arguments, one can write the contribution due to the Landau polaron-polaron interaction by adding to Eq.~\eqref{RFShift3} the following term:
\beq
\Delta_\text{int} = {6\over 5}\, \alpha\,\varepsilon_F\, F(\kappa_Fa) \,\bar x
  \label{eq:F-RF}
\eeq
where $\alpha=(\bar n_2/\bar n_1)/\bar x$ is a factor smaller than 1. Here, $\bar x$ and $\bar n_i$ are the concentration and the density of state-\ket{i} atoms averaged over the integration region.
In our temperature and concentration regime ($\bar x < 0.5$), we find $\alpha$ to significantly reduce the effect of the polaron interaction with respect to the naive expectation $\varepsilon_F\,F(\kappa_Fa)\,\bar x $. Although we refrain from providing a quantitative estimate, such an effect could motivate the absence of polaron-polaron interaction effects in our measurements.
 
\vspace{5mm}
\noindent\textbf{Initial-state interactions}\\[2mm]
In the region of parameters considered in our experiment, state-\2 impurities experience a weak repulsion with the state-\1 Fermi gas and therefore form themselves repulsive quasiparticles, so that their energy and effective mass differ slightly from the bare-particle ones.
The data presented in Fig.~3 of the main text are extracted taking into account the initial-state interactions by replacing Eq.~\eqref{RFShift3} with the following expression:
\beq
\Delta_\pm = E_{\pm}(\kappa_F a_{13}) - E_{+}(\kappa_F a_{12}) + \bar \varepsilon \left({m\over m_{\pm}^*(\kappa_F a_{13})} - {m\over m_{+}^*(\kappa_F a_{12})} \right)\:.
\eeq
The initial-state interactions are however very weak in the whole regime probed, $0.07 < \kappa_F a_{12} < 0.17$, so that the contributions of $E_{+}(\kappa_F a_{12})$ and $m_{+}^*(\kappa_F a_{12})$ may be safely estimated by means of the second-order formula given in Eq.~\eqref{RFshiftLHY}.
Namely, for all investigated interaction strengths we have $0.03 < E_{+}(\kappa_F a_{12}) < 0.08\,\varepsilon_F$ and $0.99<m/m_{+}^*(\kappa_F a_{12}) <1$.

\subsection{Rabi oscillations}
In our experiment, we probe the coherence properties of the repulsive and attractive polarons by driving Rabi oscillations between the atomic states $|2\rangle$ and $|3\rangle$.
Linear response theory predicts that the impurities will perform coherent Rabi oscillations at a frequency $\Omega$ equal to the bare Rabi frequency $\Omega_0$ times the overlap between the initial and final states,
\beq
\Omega =\Omega_0\langle\psi_3|\psi_2\rangle.
\eeq
For a non-interacting initial state, this formula simply reduces to $|\Omega /\Omega_0|=\sqrt{Z}$ \cite{Kohstall2012}.

In the case of the \Li mixture here investigated, however, interactions affect not only the final state $|3\rangle$, but also the initial state $|2\rangle$, though more weakly. In this more complicated case, the overlap can still be computed, by means of the single particle-hole Ansatz (see Eq.~\eqref{1PH_ansatz}).
Using the explicit expression for the particle-hole terms 
\beq\label{amplitudes_phi}
\qquad \phi_{\bq\bk}=\sqrt{Z}\,\frac{T(\bp+\bq,E_\down+\xi_{\bq\up})}{E_\down-E_{\bp\bk\bq}},
\eeq
we can write the overlap as 
\begin{align}
\langle\psi_3|\psi_2\rangle = \sqrt{Z_2 Z_3}\left(1-  \sum_{|q|<k_F} \;T^*(\bp+\bq,E_{\down 3}+\xi_{\bq\up})T(\bp+\bq,E_{\down 2}+\xi_{\bq\up})\frac{\Pi^*(\bp+\bq,E_{\down 3}+\xi_\bq)-\Pi(\bp+\bq,E_{\down 2}+\xi_\bq)}{{E_{\down3}^*-E_{\down2}}}\right), \rule[-2.5em]{0pt}{0pt}
\end{align}
with the retarded pair propagator $\Pi(\bp+\bq,E_\down+\xi_q)=\sum_{\bk}\left(\frac {{\theta(k-k_F)}}{E_{\down3}-E_{\bp\bk\bq}+i0_+}+\frac{2m_r}{k^2}\right)$. The symbols $\Pi^*$ and $T^*$ denote instead the advanced pair propagator and T-matrix, respectively, while $Z_2$ and $Z_3$ denote the residues of state-\2 and \3 quasiparticles.
This expression reduces correctly to two known limits. In particular when $a_{12}\rightarrow 0$ one gets the simple result $|\Omega/\Omega_0|=\sqrt{Z_3}$ \cite{Kohstall2012}. For $a_{12}\rightarrow a_{13}$, instead, using the definition of the residue $Z$ one obtains that $|\Omega/\Omega_0|=1$. The complete expression is used to compute the solid curves in Fig.~4b of the main text. Our repulsive polaron Rabi oscillation data seem also to agree well with recent predictions on impurity coherent dynamics in Ref.~\cite{Parish2016}.

\section{Attractive polaron measurements}
\subsection{Radio-frequency spectral response}

\begin{figure}[t!]
\vspace*{8mm}
\begin{center}
\includegraphics[width=\columnwidth]{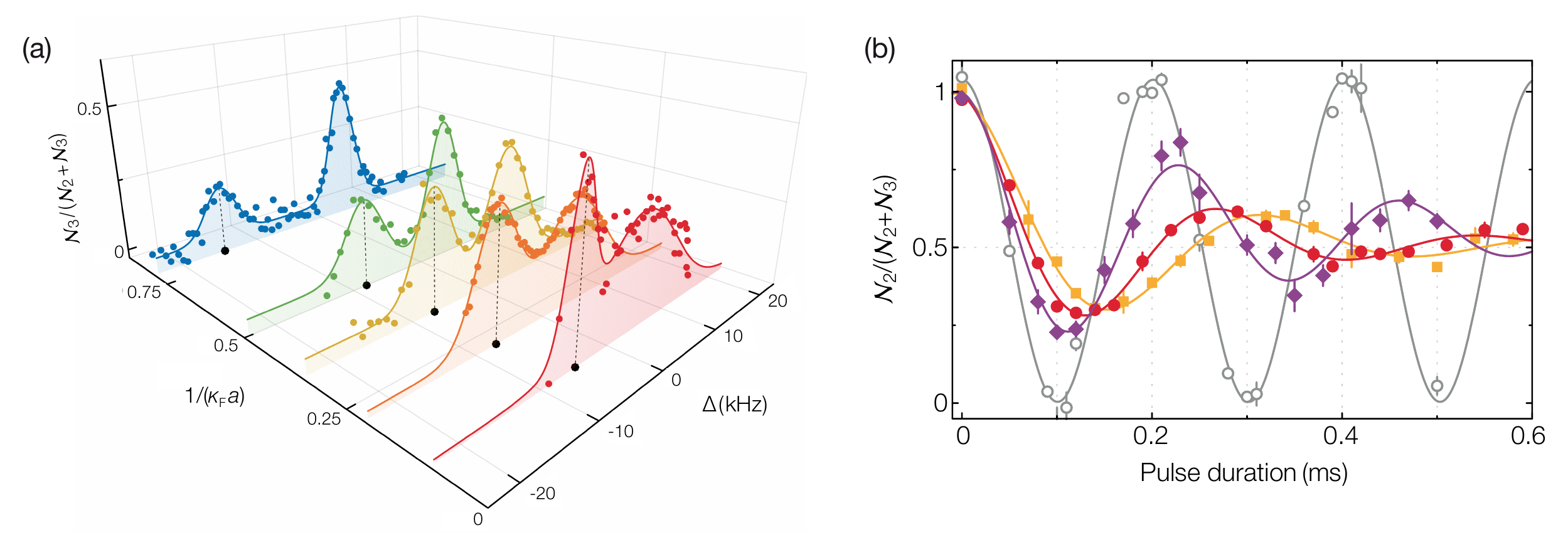}
\caption{(a) High RF power spectra recorded at various $1/(\kappa_F a)$ values with impurity concentration $x=0.15(3)$. The experimentally recorded spectra are shown together with double-Gaussian fits used to extract the resonance positions. The attractive polaron resonance is up-shifting in frequency for decreasing $1/(\kappa_F a)$ towards unitarity. (b) Attractive polaron Rabi oscillations at $x=0.15(3)$ for $1/(\kappa_F a) = 0.5$ (yellow squares), 0.25 (red circles) and 0 (purple diamonds). Rabi $\2 \leftrightarrow \3$ oscillations in a non-interacting gas are shown for comparison (empty grey circles). Solid lines are the damped sinusoidal fits used to extract $\Omega$. Error bars denote the s.e.m. of experimental data.}
\label{FigS3}
\end{center}
\end{figure}

While in the main text we focus the discussion on the repulsive polaron energy, our spectroscopic scheme allows to obtain information about the energy of the attractive polaron as well (see inset of Fig. 3a in the main text). Typical spectra displaying both attractive and repulsive polaron resonances are shown in Fig.~\ref{FigS3}, together with double-Gaussian fits used to extract the resonance positions. To increase the attractive polaron signal strength, we recorded spectra at strong coupling $1/(\kappa_F a) < 0.8$ using a higher RF power with respect to the one used for the repulsive polaron spectroscopy presented in Fig.~2a of the main text. In particular, the spectra presented in Fig.~\ref{FigS3} are recorded using 0.5\,ms-long pulses with RF power increased by 20\,dB, resulting in a pulse area of 5$\pi$ for the non-interacting impurity gas.
As briefly mentioned in the main text, the attractive polaron signal nearly overlaps with the one from molecular excitations. For this reason we are not able to extract $E_-$ and $m^*$ through a systematic study of the spectra at varying concentration $x$. This is further complicated by the attractive polaron dispersion, that for $1/(\kappa_F a) \in [0, 0.8]$ does not deviate substantially from the bare-particle one, making the dependence of $\Delta_-$ on $x$ intrinsically weaker than that expected for the repulsive polaron. Therefore, we assume the mean value of $\Delta_-$  measured for $x$ between 0.05 and 0.3 to be a good estimate of $E_-$ for all interactions (and taking into account the initial state 1-2 interactions). The results (see inset of Fig.~3 in the main text) are indeed found in good agreement with theory \cite{Prokofev2008, Schirotzek2009, Schmidt2011, Massignan2011, Vlietinck2013} and previous experimental studies performed by means of direct, spatially-resolved RF spectroscopy on a polarized Fermi gas \cite{Schirotzek2009}.

\subsection{Attractive polaron Rabi oscillations}
In order to extract the residue $Z$ of the attractive polaron, we perform Rabi oscillations at the attractive polaron resonance for various interaction strengths $1/(\kappa_F a)$. Sample Rabi oscillations are displayed in Fig.~\ref{FigS3}. In contrast to the case of the repulsive polaron, only a renormalized frequency and a damping of the oscillation amplitude are necessary to fit the data with excellent agreement: since no decay to lower-lying states is possible, no decay of the oscillation offset is observed and the relative population tends to saturate at 0.5 as expected for pure decoherence processes.

\section{Collision-induced decoherence}\label{SecDecoh}
Besides an energy shift, interactions with the surrounding medium cause a collisional broadening of the RF spectra. Similarly, impurity-bath interactions do not only cause a renormalization of the Rabi frequency, but they also introduce a damping of the Rabi oscillations.
This effects can be intuitively understood in the framework of the impact theory of pressure-induced effects on spectral lines \cite{Sobelman1972}, which assumes the collisions of the impurity atoms with the majority particles to be effectively instantaneous.
Given an impurity-fermion forward scattering amplitude $f(\epsilon)$ for a given collision energy $\epsilon$, this approach predicts Lorentzian profiles, whose line shift (width) is proportional to the real (imaginary) part of the forward scattering amplitude $\langle f(\epsilon)\rangle$, averaged over all possible collision energies.
The optical theorem relates the imaginary part of $\langle f(\epsilon)\rangle$ to the average elastic scattering rate $\gamma \propto \text{Im}(\langle f(\epsilon)\rangle)$. This results in a finite $1/\gamma$ lifetime for the coherence of the impurity wavepacket, and causes Lorentzian broadening with a full width at half maximum (FWHM) $\gamma/(2 \pi)$.
Moreover, elastic scattering processes randomly change the impurity momentum, hence the phase of the associated wavepacket. This causes collisional decoherence, i.e. a decay of the coherences between the weakly-interacting and the resonant impurity states. This leads to a damping $\gamma_R \sim \gamma$ of the Rabi oscillations \cite{Kohstall2012}, and to the decay of the contrast of Ramsey fringes in spin-echo experiments \cite{Cetina2015, Cetina2016}.
It is important to stress that such a dephasing markedly differs from inelastic decay, which affects the impurity state populations. As a consequence, one generally expects the collisional damping rate $\gamma$ to differ from the population decay rate $\Gamma$.

\begin{figure}[t!]
\begin{center}
\includegraphics[width= 8.6 cm]{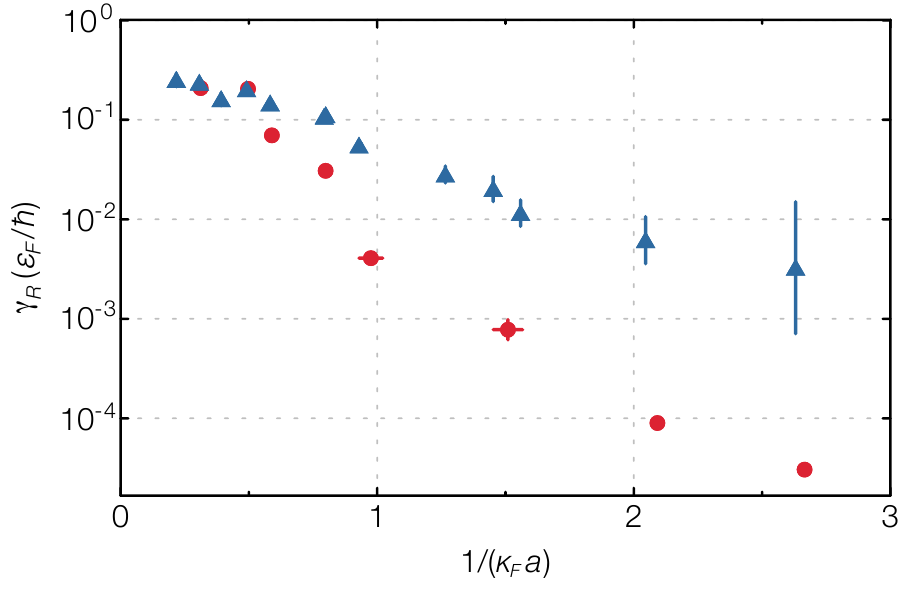}
\caption{Damping rate $\gamma_R$ of the repulsive polaron Rabi oscillations (blue triangles). The data points for $\gamma_R$ are obtained by fitting the repulsive polaron Rabi oscillations with the function given in Section~\ref{RabisSec}.
The repulsive polaron population decay rate $\Gamma$ (red circles) measured through a double $\pi$-pulse sequence (see Section \ref{DecaySec}) is also shown for comparison (see also Fig.~4 of the main text).}
\label{FigS4}
\end{center}
\end{figure}

\begin{figure}[b!]
\begin{center}
\includegraphics[width= \columnwidth]{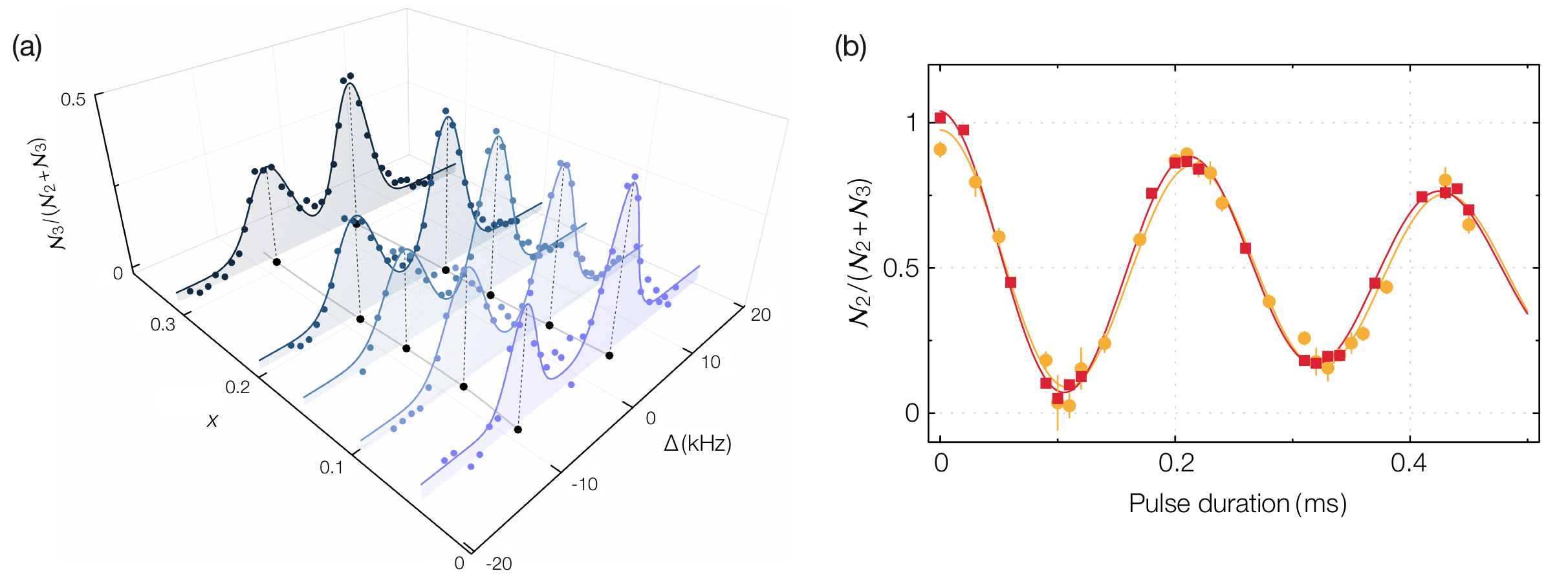}
\caption{(a) Polaron spectra at $1/(\kappa_F a) \simeq 0.5$ for various value of the relative concentration $x$. Double-Gaussian fits are also shown as solid lines. No increase of the repulsive polaron spectral width is detected upon increasing $x$. (b) Repulsive polaron Rabi oscillations at $1/(\kappa_F a) \simeq 1$ for $x \simeq 0.05$ (yellow circles) and $x\simeq0.31$ (red squares). Solid lines are the damped sinusoidal fits, while error bars denote the s.e.m. of experimental data.}
\label{FigS5}
\end{center}
\end{figure}

\enlargethispage{11pt}
Similarly to what already observed for ${}^{40}$K impurities in a \Li Fermi gas, our RF spectroscopy and Rabi oscillation experiments are strongly affected by collisional broadening and collision-induced decoherence.
In contrast with the textbook case of a two-level system described by the optical Bloch equations for which $\gamma = \Gamma/2$, our strongly interacting impurities experience a collisionally opaque medium, which causes $\gamma$ to greatly exceed $\Gamma$ \cite{Sobelman1972} (see Fig.~\ref{FigS4}).
This is especially evident if attractive, rather than repulsive polaron Rabi oscillations are considered (see Fig.~\ref{FigS3}b): for the attractive branch in the regime herein investigated $\Gamma=0$ since the attractive polaron is the ground state of the many-body system. Still, we observe damping rates of the Rabi oscillations comparable to the ones observed for the repulsive case at similar values of $1/\kappa_F a$.

\vspace{5mm}
\noindent\textbf{Polaron-polaron interactions}\\[2mm]
While a detailed characterization of the decoherence rates of the polaronic branches is beyond the scope of this work, a closer look to both the spectral widths and Rabi oscillation damping rates provides additional important information associated with possible polaron-polaron interaction effects.
If quasiparticle effective interactions played a significant role in our system, they would appear as an additional source of decoherence and broadening, which should increase for increasing impurity concentration \cite{Cetina2016}.
In Fig.~\ref{FigS5} we show that the repulsive polaron spectral width does not exhibit any appreciable dependence upon $x$, although $\Delta_+$ increases while decreasing $x$ and the spectral gap between the attractive and repulsive polaron consequently increases. This trend is not limited to a specific $\kappa_F a$ value, but is generic for all interaction regimes we explored. Similarly, neither the damping rate nor the frequency of Rabi oscillations present a detectable dependence on concentration. As an example, in Fig.~\ref{FigS5} we compare data for the repulsive polaron branch at $\kappa_F a\simeq 1$ for $x\simeq0.05$ and 0.31, which overlap within the experimental uncertainty.
We therefore conclude that quasiparticle interaction effects are either below the experimental resolution of our spectroscopic studies.


\end{document}